\documentclass[fleqn,usenatbib,useAMS,linenumbers]{mnras}

\usepackage{graphicx}	% Including figure files
\usepackage{scalerel}

\usepackage{amsmath}	% Advanced maths commands
\usepackage{amssymb}	% Extra maths symbols

\usepackage{multicol}        % Multi-column entries in tables
\usepackage{bm}		% Bold maths symbols, including upright Greek
\usepackage{pdflscape}	% Landscape pages
\usepackage[T1]{fontenc}
\usepackage{ae,aecompl}

\usepackage{newtxtext,newtxmath}
\usepackage{kotex}
\usepackage{threeparttable}
\usepackage{amssymb}
\usepackage{hyperref}
\usepackage[tableposition=top]{caption}

\title[RPS and ISM disc Truncation]{Ram Pressure Stripping and ISM disc Truncation\\: Prediction vs. Observation}

\author[S. Lee et al.]{Seona Lee$^{1}$, Yun-Kyeong Sheen$^{2}$, Hyein Yoon$^{1,3,4}$, Yara Jaff\'e$^{5}$ and Aeree Chung$^{1}$
\\
\\
$^{1}$Department of Astronomy, Yonsei University, 50 Yonsei-ro, Seodaemun-gu, Seoul 03722, Korea
\\ $^{2}$Korea Astronomy and Space Science Institute, 776 Daedeok-daero, Yuseong-gu, Daejeon 34055, Republic of Korea
\\ $^{3}$Sydney Institute for Astronomy, School of Physics, A28, The University of Sydney, Sydney, NSW 2006, Australia
\\ $^{4}$ARC Centre of Excellence for All Sky Astrophysics in 3 Dimensions (ASTRO 3D)
\\ $^{5}$Instituto de F\'isica y Astronom\'ia, Facultad de Ciencias, Universidad de Valpara\'iso, Avda. Gran Bretana 1111 Valpara\'iso, Chile}

\date{Last updated 2022 September 30}

\pubyear{2022}

\begin{document}
\label{firstpage}
\pagerange{\pageref{firstpage}--\pageref{lastpage}}
\maketitle

\begin{abstract}

Ram pressure stripping (RPS) is known to be a key environmental effect that can remove interstellar gas from galaxies in a cluster. The RPS process is commonly described as a competition between the ram pressure by the intracluster medium (ICM) and the anchoring pressure on the interstellar medium (ISM) by the gravitational potential of a galaxy. However, the actual gas stripping process can be more complicated due to the complexity of gas physics such as compression and geometrical self-shielding as well as cooling and heating. In order to verify how well the observed signatures of the RPS process can be understood as simple momentum transfer, we compare the stripping radii of Virgo cluster galaxies in different stages of RPS measured from the \ion{H}{i} observation with the predicted gas truncation radii for the given conditions. For the sample undergoing active RPS, we generally find good agreements between predictions and observations within a measurement uncertainty.
On the other hand, galaxies likely in the early or later RPS stage and/or the ones with signs of environmental impacts other than RPS such as tidal interaction or starvation, show some  discrepancies. 
Our results imply that the conventional RPS relation works reasonably well in a broad sense when RPS is the most dominant process and the galaxy is located where the surrounding environment can be well defined.
Otherwise, more careful inspections on the second mechanism and local environment are required to assess the impact of RPS on the target.

\end{abstract}

%%%%%%%%%%%%%%%%%%%%%%%%%%%%%%%%%%%%%%%%%%%%%%%%%%

%%%%%%%%%%%%%%%%% BODY OF PAPER %%%%%%%%%%%%%%%%%%

% The MNRAS class isn't designed to include a table of contents, but for this document one is useful.
% I therefore have to do some kludging to make it work without masses of blank space.
%\begingroup
%\let\clearpage\relax
%\tableofcontents
%\endgroup
%\newpage

\section{Introduction}

The cluster environment is considered as one of the key drivers for galaxy evolution.
The mean characteristics of cluster galaxies are clearly distinct from those in low-density environments in that early-type galaxies are more populated and star formation is generally less active with lower cool gas fraction \citep[see][and references therein]{vanGorkom2004,Boselli2006,Park2009,Cortese2021,Boselli2022}.
This implies that dense environments can make significant impacts on the properties of galaxies.

Among many environmental processes, ram pressure due to the intracluster medium (ICM) is found to be the mechanism which affects the interstellar medium (ISM) most severely without disturbing the stellar component of galaxies \citep{Chung2009}.
According to \citet{Gunn1972}, galaxies may lose the ISM while falling into a cluster when the ram pressure from the ICM exceeds the anchoring pressure by the potential of a galaxy, i.e.,
\begin{equation}
    \rho_{\scaleto{\rm ICM}{4pt}} v_{\rm rel}^2 > 2\pi G \Sigma_{\rm g}(r) \Sigma_{\rm \star}(r)
    \label{eq:GG}
\end{equation}
where $\rho_{\scaleto{\rm ICM}{4pt}}$ is the ICM density, $v_{\rm rel}$ is the velocity of the galaxy with respect to the ICM, and $\Sigma_{\rm g}$ and $\Sigma_{\rm \star}$ are the gas and stellar surface density of the galaxy, respectively.
Based on this relation, the ram pressure can remove the ISM from a galaxy outside the radius at which the ram pressure balances with the anchoring pressure. This is a simple model that assumes for instantaneous stripping from an infinitely thin disc falling face-on into the ICM.

Currently, our understanding of ram pressure stripping largely relies on equation (\ref{eq:GG}), i.e., the ISM stripping condition based on Gunn \& Gott (hereafter, GG)'s formula.
From the observational perspective, this RPS equation is commonly used to make sense out of the observed gas deficiencies and the truncation of the gas disc \citep[e.g.][]{Kenney2004,Chung2007}.
Also in the RPS models,  GG's stripping condition is widely adopted regardless of the simulation tools (e.g. SPH, smoothed particle hydrodynamics, AMR, etc) and/or initial conditions \citep[e.g.][]{Abadi1999,Vollmer2001,Roediger2005, Roediger2006,Kronberger2008,Tonnesen2009,Ramos-Martinez2018,Lee2020}.

In practice, however, RPS is likely to be a complex process.
The physical conditions of the ISM can be changed while being pushed by the ram pressure.
As a galaxy starts to fall into a cluster, before experiencing gas stripping, its gas density can increase towards the leading side of the galaxy through gas compression.
This has been seen in both observations and simulations \citep{Vollmer2001,Kenney2004,Chung2007,Vollmer2006,Tonnesen2009,Tonnesen2021}.
This also may cause an excess of molecular gas \citep{Vollmer2008,Henderson2016,Lee2017,Moretti2018,Moretti2020,Cramer2021} and potentially the enhancement of the star formation \citep{Fujita1999,Koopmann2004b,Kronberger2008,Boselli2014,Vulcani2018,Lizee2021}, which will change both the gas and stellar surface densities of equation (\ref{eq:GG}).
In dwarf galaxies, RPS can also cause stellar warps in the opposite direction to the stripped tails \citep{Smith2013}.
The relative velocity with respect to the ICM as well as the surrounding ICM density are also likely to keep varying as galaxies move around, and the ICM is not always smooth and static \citep{Shibata2001}.

The limitations in the understanding of RPS based on GG's description have been raised in previous studies.
For instance, \citet{Jachym2007} pointed out that the duration of the interaction between ICM and ISM, and/or the encounter angle of the galactic disc with the ICM should be carefully taken into account as those can be critical to understand the target based on this RPS condition.
\citet{Steinhauser2016} warned that the actual stripping radius can be larger than analytic estimates due to the redistributed gas after the peak of RPS.

Hence there have been some efforts to verify the scope which the RPS formula can be applied to.
\citet{Gullieuszik2020} measured the truncation radius of the ionized gas (H$\alpha$) in a sample of stripped "jellyfish" galaxies observed with MUSE by the GASP survey \citep{Poggianti2017} and compared it with the predicted gas truncation radius based on equation (\ref{eq:GG}), finding that they match reasonably well (with some scatter, see their fig. 7).
The short-coming of using H$\alpha$ is that its characteristics may keep changing due to the in-situ star formation triggered by ram pressure, and so it may not be the best tool to verify equation (\ref{eq:GG}). Instead, \ion{H}{i} is more straightforward as it is less likely to get directly involved with star-forming activity. It is also usually the most extended and hence loosely bound galactic component, and it makes the best tracer of environmental processes including RPS. \citet{Koppen2018} calculated the maximum ram pressure that one galaxy might have experienced and compared it with the local ram pressure, showing that equation (\ref{eq:GG}) works quite well for the objects close to be at the cluster centre. \citet{Koppen2018} however used the azimuthally averaged \ion{H}{i} radius, which does not necessarily represent the true gas truncation especially for the cases undergoing active RPS, thus showing highly asymmetric gas morphology.
Furthermore, most previous studies including \citet{Koppen2018} have adopted the gas surface density model before stripping in their analysis although the density is likely to be decreased during the stripping process. 
These might have resulted in the overestimation of the ram pressure.

The main goal of this study is to re-assess the applicability of GG's relation by comparing the prediction based on the equation with the observational measurement. The question is whether we find general agreements between theory and observation, and also with the results from other similar works but based on different ram pressure and/or galaxy parameters from ours such as \citet{Koppen2018}. By addressing the answers to these questions, this work aims to establish how broadly GG’s relation can be used to understand the snap shots of individual galaxies and their evolutionary histories. This paper is organized as follows. In Section 2, we introduce the sample. In Section 3, we describe how \ion{H}{i} extents are estimated using equation (\ref{eq:GG}), and how truncation radii are observationally measured. In Section 4, we compare the predictions and the observations from the previous section. We also compare our results with those of \citet{Koppen2018} which used the same \ion{H}{i} data. In Section 5, we discuss the scope of GG's formula to which RPS can be applied and its limitations. In Section 6, we summarize and conclude.
Throughout this paper, we adopt a distance of 16.5~Mpc for the Virgo cluster \citep{Mei2007}.

\begin{figure}
 \includegraphics[width=\columnwidth]{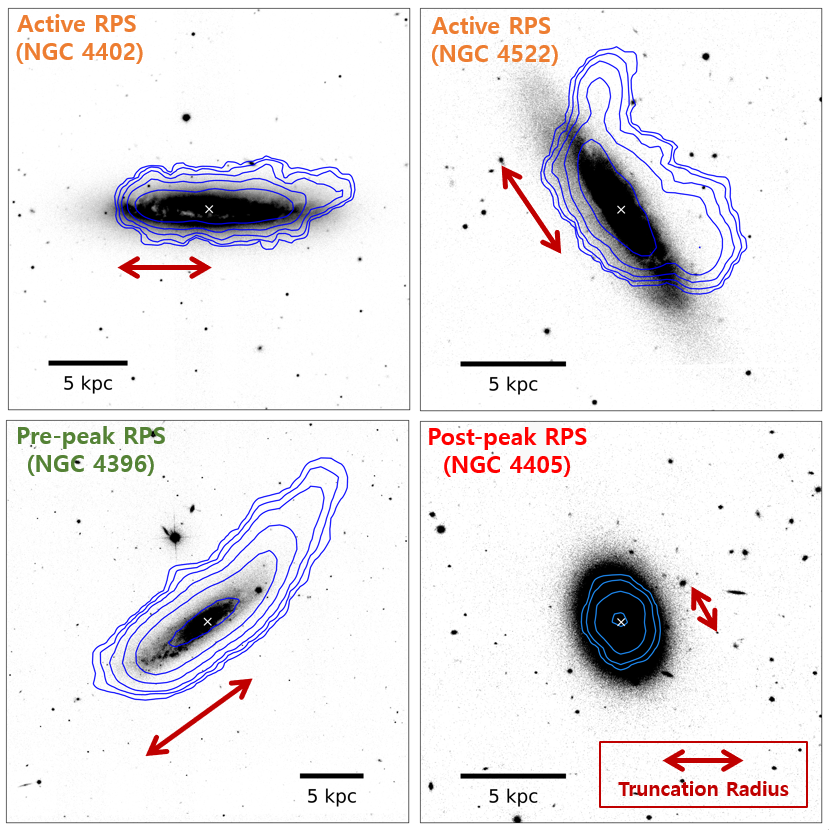}
 \caption{Concept of truncation radius, $R_{\rm t}$. In each panel, the SDSS $\it {i-}$band image is shown in grey scale, the \ion{H}{i} distribution is shown by blue contours, and the galactic centre in optical is marked as cross. A typical resolution of VIVA \ion{H}{i} images is 15$\arcsec$ or 1.2 kpc at a Virgo distance of 16.5 Mpc. In this work, as indicated by a red line, $R_{\rm t}$ is defined at the radius where the \ion{H}{i} surface density drops to $1~M_{\odot}~{\rm pc}^{-2}$ regardless of whether \ion{H}{i} is stripped within the extent of the stellar disc (active and past RPS cases) or not (pre-peak RPS cases). In these examples, $R_{\rm t}$ is shown only for one direction per target but $R_{\rm t}$ has been measured in eight directions for each galaxy to take into account the asymmetry in \ion{H}{i}. Further details can be found in Section \ref{chap:method}.}
 \label{fig:Schematic}
\end{figure}

\begin{figure*}
 \centering
 \includegraphics[width=2\columnwidth]{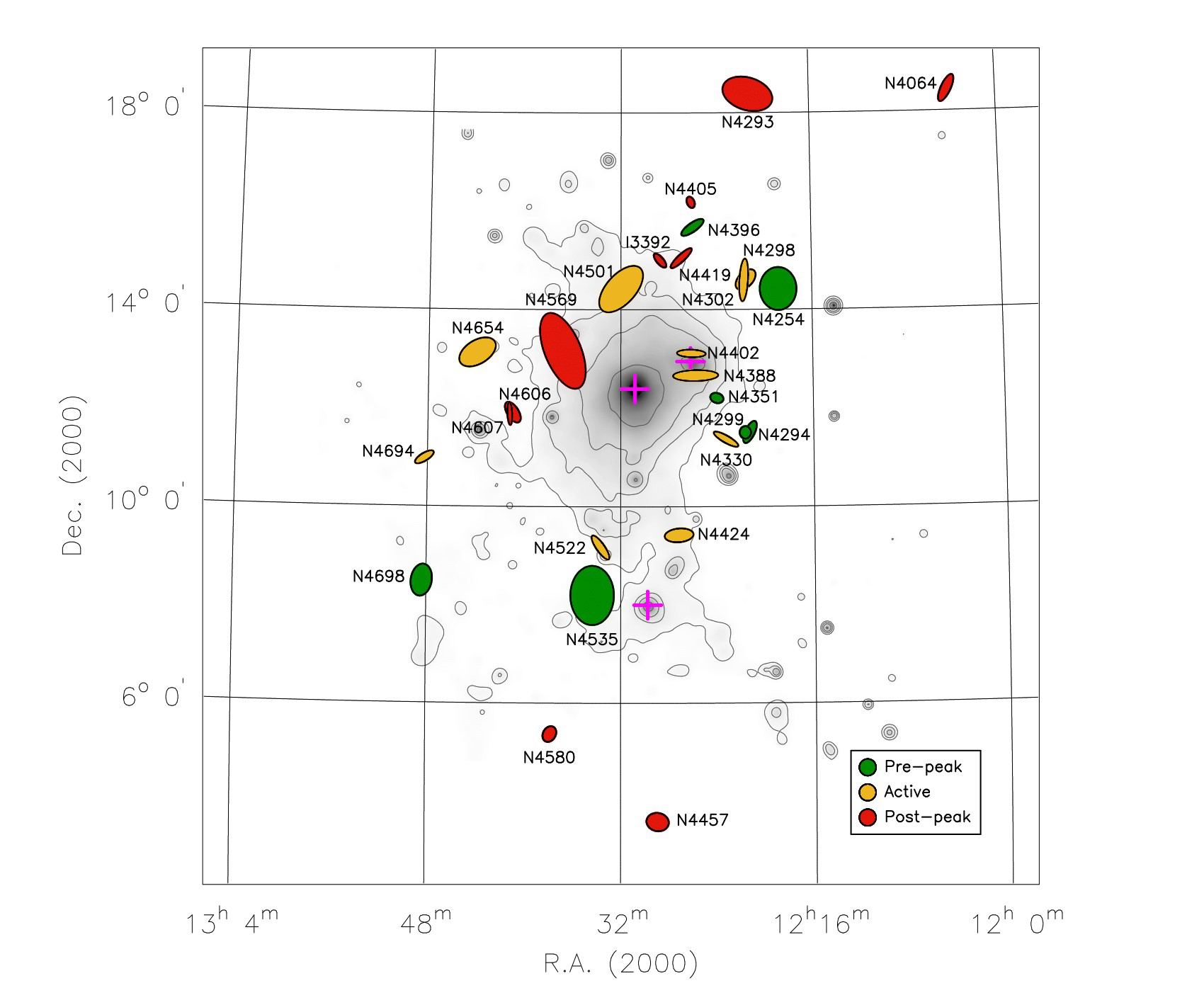}
 \caption{The locations of the sample are shown overlaid on the X-ray image \citep{Bohringer1994} of greyscale and grey contours. Ellipses represent the optical size~$\times$~10, the inclination and the position angle of individual galaxies which are colour coded by the ram pressure stripping class of \citet{Yoon2017}. Magenta crosses indicate the centres of three major sub-clusters, i.e., M86, M87, and M49 (from north to south).}
 \label{fig:xray}
\end{figure*}

\section{Observational Data} \label{chap:data}

In this work, we make use of the data from the VLA Imaging survey of Virgo galaxies in Atomic gas \citep[VIVA,][]{Chung2009}. As the nearest rich cluster, Virgo is an ideal laboratory to study the environmental effects on galaxy evolution in great detail. The VIVA study provides high-resolution \ion{H}{i} images for 52 late-type galaxies with a typical resolution of 15\arcsec and a column density sensitivity of $\rm 3-5\times10^{19}\,cm^{-2}\,in\,3\sigma\,per\,\approx 15\arcsec\,beam\,and\,10\,km\,s^{-1}\,channel\,width$. Its superior spatial resolution and sensitivity help to reveal many galaxy samples undergoing various processes including RPS.
Besides, RPS is a rapid process compared to other gas stripping mechanisms, allowing a single snapshot of \ion{H}{i} image to best provide the current condition of RPS including the size of the truncated disc in this analysis.
The VIVA galaxies have been well studied through their \ion{H}{i} properties in \citet{Chung2007,Chung2009} and phase-space diagrams in \citet{Yoon2017}. Combining the information from those previous studies with a quantitative approach using the GG's equation will lead to a deeper understanding of the samples.

For our study, the galaxies showing the signs of active RPS with no strong evidence for any other environmental effects such as tidal interactions make the most suitable targets. 
The clearest cases would be the ones that have a highly asymmetric gas disc. They all show \ion{H}{i} truncation within the extent of the relatively unperturbed stellar disc at least in one direction \citep{Kenney2004,Crowl2005,Chung2007}. The examples of active RPS are also normally found in the locations where measurable X-ray background emission is present from which the strength of ram pressure can be constrained. Hence the GG's gas stripping condition can be tested for this group of galaxies with less complexity compared to the cases showing signs of other processes than RPS and/or no detected surrounding X-ray emission.

The VIVA sample has been classified as four RPS stages by \citet{Yoon2017}.
Class 0 is galaxies with no signs of RPS in \ion{H}{i} morphology and deficiency, whereas Class I to III are those that are observed to be experiencing ram pressure. Class I shows an asymmetric \ion{H}{i} disc with mild compression on one side yet no sign of truncation within the extent of the stellar disc. Together with extended gas tail features often found among them, Class I makes candidates for galaxies at early RPS stages during the first infall to the cluster. Class II galaxies are the ones with a highly asymmetric \ion{H}{i} disc. Some of Class II also look tidally disturbed, yet they all show clear \ion{H}{i} truncation at least in one direction. Class II galaxies are therefore the best candidates that are likely under the influence of strong ram pressure. Lastly, Class III is galaxies with a relatively symmetric yet severely truncated \ion{H}{i} disc, making good candidates for galaxies which have already lost significant amounts of gas via RPS. Some Class III galaxies are found in rather low-density environments at large clusto-centric distances, which are thought to be the examples of back-splashed cases. More quantitative criteria for different classes can be found in \citet{Yoon2017}. There is another group (Class IV) which is more likely to be affected by other mechanisms than ram pressure, and hence this group is not included in this study. The locations of the studied galaxies are shown in Fig. \ref{fig:xray} overlaid on the X-ray image.

In addition to the \ion{H}{i} data from the VIVA, we use the optical $\it {i-}$band sky-subtracted images from the Sloan Digital Sky Survey (SDSS) DR12\footnote{https://www.sdss.org/dr12/} \citep{York2000,Eisenstein2011}. More details on optical data will be described in Section \ref{sec:AP}.

\section{Measuring Truncation Radii} \label{chap:method}

This section describes how we calculate a predicted \ion{H}{i} extent using GG's equation for a given stellar mass at a given location, and how we measure an \ion{H}{i} extent from the VIVA observational data. In this work, the truncation radius ($R_{\rm t}$) is simply defined at the radius where the \ion{H}{i} surface density drops to $1~M_{\odot}~{\rm pc}^{-2}$ as shown in Fig. \ref{fig:Schematic}. To consider the gas asymmetry, $R_{\rm t}$ has been measured in eight different directions for all sample galaxies, yielding eight $R_{\sc t}$ for each galaxy. The truncation radius generally refers to the gas extent when the gas is stripped within a stellar disc. Strictly speaking, the \ion{H}{i} truncation radius usually refers to the \ion{H}{i} extent for such a case where \ion{H}{i} gas is likely to have been stripped and hence less extended compared with the stars. However, in this study, assuming that all the galaxies in the sample are undergoing ram pressure stripping to some degree, we use the terminology of truncation radius to simply refer to the \ion{H}{i} extent regardless whether they are truncated in \ion{H}{i} or not yet. This is not only for convenience but also to emphasize the fact that the main targets are the Class II galaxies, particularly to the directions with clearly visible \ion{H}{i} truncation.

\subsection{Predicted $R_{\rm t}$ from Gunn \& Gott's equation}
\label{sec:theory}

\subsubsection{Ram pressure}
\label{sec:RP}

The ram pressure $(P_{\rm ram})$ is proportional to the ICM density and the square of the galaxy's velocity relative to the ICM. In addition, the encounter angle between the ICM and the galactic disc is also important to estimate the effective $P_{\rm ram}$. That is, the face-on encounter (i.e., the wind pushes the disc plane perpendicularly) makes a much greater impact compared to the edge-on encounter (i.e., the wind pushes the disc from the side), and the ISM is stripped more efficiently \citep{Roediger2006,Jachym2009}. Based on these, the effective $P_{\rm ram}$ can be expressed as follows:
\begin{equation}
\label{eq:P_ram}
P_{\rm ram} = \cos^2{\phi}\rho_{\scaleto{\rm ICM}{4pt}} v_{\rm rel}^2
\end{equation}
where $\phi$ is the angle between the galactic rotation axis and the ICM, $\rho_{\scaleto{\rm ICM}{4pt}}$ is the ICM volume density, and $v_{\rm rel}$ is the galaxy's velocity relative to the ICM.

\begin{enumerate}
    \item ICM density
    
    We estimate the ICM density $(\rho_{\scaleto{\rm ICM}{4pt}})$ using the standard $\beta$-model \citep{Cavaliere1976}. In the Virgo cluster, the X-ray peak coincides with the centre of M87. Thus taking the location of M87 as the cluster centre, the ICM density of the Virgo cluster can be described as follows:
    \begin{equation}
        \rho_{\scaleto{\rm ICM}{4pt}} (d_{\rm M87,3D})=\frac{\rho_{0,\rm ICM}}{(1+d_{\rm M87,3D}^2/d_{\rm c}^2)^{\frac{3}{2}\beta}}
    \end{equation}
    where $d_{\rm M87,3D}$ is the three-dimensional distance between the cluster centre and each galaxy, $d_{\rm c}$ is the core radius of the cluster, and $\rho_{0,\rm ICM}$ is the ICM density at the cluster centre. The best-fitting values from the ROSAT X-ray observation are $\rho_{0,\rm ICM}=4.2\times10^{-2} cm^{-3}$, $\beta=\frac{7}{15}$, and $d_{\rm c}=2.7\arcmin$ or 13 kpc at the distance of the Virgo cluster \citep[=16.5 Mpc;][]{Mei2007} \citep{Schindler1999, Vollmer2009}.
    Assuming spherical symmetry, we convert the projected clusto-centric distance $d_{\rm M87}$ into a three-dimensional distance by averaging the projection angle, $d_{\rm M87,3D}=\frac{\pi}{2}d_{\rm M87}$.
    
    When compared with the X-ray observation, the $\beta$-model-based ICM density can differ by up to a factor of five. However, it only causes 1$-$2 kpc of discrepancies in the prediction of $R_{\rm t}$, which is comparable to the observational measurement uncertainty.\\
    
    \item Relative velocity
    
    We assume that the velocity of the target galaxy relative to the ICM ($v_{\rm rel}$) ranges between the Keplerian velocity $(v_{\rm \scaleto{\rm Kep}{6pt}})$ and the escape velocity $(v_{\rm esc})$ in NFW potential \citep{Navarro1996}, i.e., $v_{\rm \scaleto{\rm Kep}{6pt}} < v_{\rm rel} < v_{\rm esc}$, where $v_{\rm esc}$ is estimated following \citet{Jaffe2015} and $v_{\rm \scaleto{\rm Kep}{6pt}}$ can be calculated from $v_{\rm esc}$ as below:
    \begin{equation}
        v_{\rm esc} =
        \begin{cases}
        \sqrt{\frac{2GM_{200}}{3R_{200}}K} &\text{$r<R_{200}$}\\
        \sqrt{\frac{2GM_{200}}{3R_{200}s}} &\text{otherwise}
        \end{cases}
    \end{equation}
    
    \begin{equation*}
        K = g_{\rm c}\left(\frac{\ln{(1+cs)}}{s}-\ln{(1+c)}\right)+1
    \end{equation*}
    \begin{equation*}
        s = \frac{d_{\rm M87,3D}}{R_{200}}
    \end{equation*}
    \begin{equation*}
        g_{\rm c} = \left[\ln{(1+c)}-\frac{c}{1+c}\right]^{-1}
    \end{equation*}
    
    \begin{equation}
        v_{\rm \scaleto{\rm Kep}{6pt}} = \frac{v_{\rm esc}}{\sqrt{2}}
    \end{equation}
    where $R_{200}=1.55 \rm~Mpc$, $M_{200}=4.2\times10^{14} M_{\odot}$, and $c$ (the concentration parameter, $R_{200}/r_{s}$) $=2.8$ for the Virgo cluster \citep{McLaughlin1999}. $d_{\rm M87,3D}$ is the same as the description in the previous section.\\
    
    \item Encounter angle
    
    The angle between the galactic disc and the ICM wind $(\phi)$ is an important factor to assess the effective $P_{\rm ram}$ \citep{Roediger2006}. However, the projection effect together with the insufficient information on the orbit makes it difficult to estimate the actual encounter angle. In this study, we therefore probe a range of $\phi$ from 0$^{\circ}$ (face-on stripping) up to 60$^{\circ}$ (nearly edge-on stripping). For the angles larger than this, the pressure is decreased by four times or more and stripping becomes quite inefficient.  
    
\end{enumerate}

\begin{figure}
 \centering
 \includegraphics[width=0.9\columnwidth]{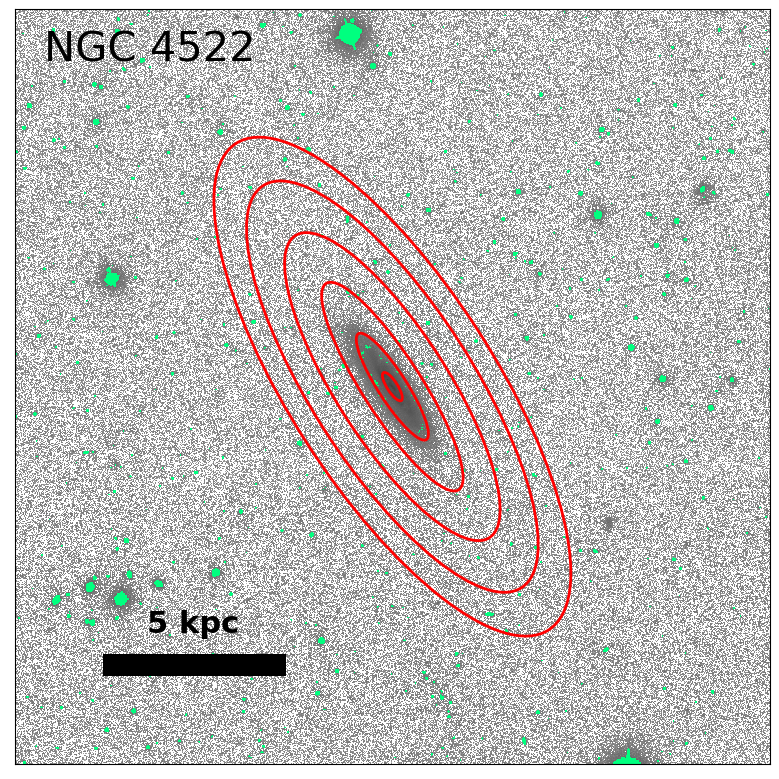}
 \caption{SDSS $\it {i-}$band image of NGC~4522. The fitted ellipses are overplotted in red on the image. The masked areas (foreground stars and other galaxies) are indicated in green.}
 \label{fig:Maskimage}
\end{figure}

\begin{figure}
 \includegraphics[width=\columnwidth]{{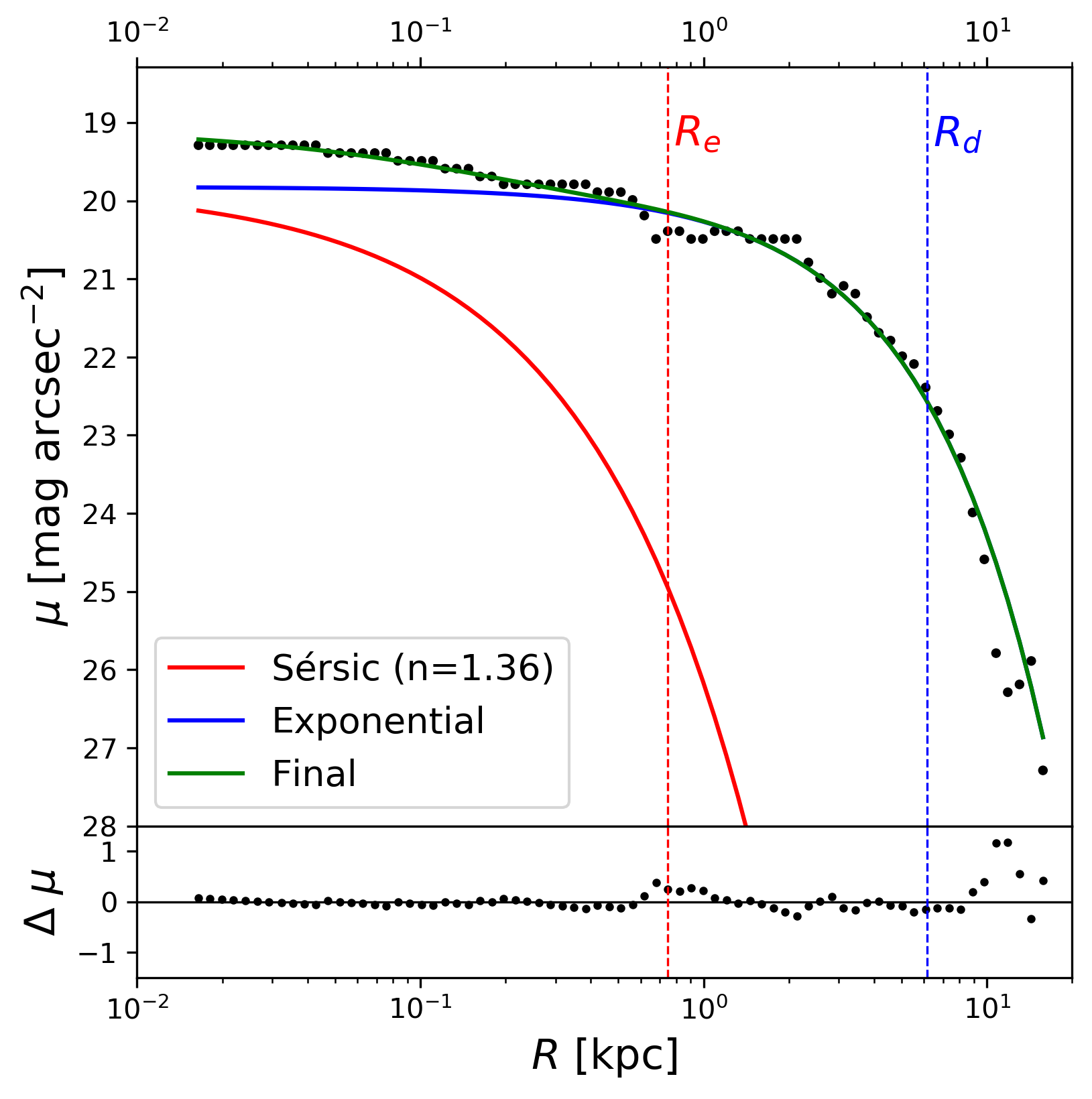}}
 \caption{Top) Stellar surface brightness profile of NGC~4522. The raw surface brightness profile is represented by black dots, the fitted/decomposed S\'ersic and exponential profiles are represented by the red and blue lines, respectively, and the total surface brightness by the green line. Bottom) The residual between the surface brightness profile from the ellipse fitting and the observed surface brightness profile.}
 \label{fig:SBimage}
\end{figure}

\subsubsection{Anchoring pressure}
\label{sec:AP}

The anchoring pressure $(P_{\rm anch})$ by the galactic potential which acts against the ram pressure can be described as follows:
\begin{equation}
    \label{eq:P_anc}
    P_{\rm anch}(r) = 2\pi G \Sigma_{\rm g}(r) \Sigma_{\rm \star}(r)
\end{equation}
where $G$ is the gravitational constant, and $\Sigma_{\rm g}$ and $\Sigma_{\rm \star}$ are the mass surface density for the ISM (\ion{H}{i} in this study) and stars. It is noteworthy that equation (\ref{eq:P_anc}) only does not include a halo component. However, the main question of this study is about whether the gas can be pushed from the disc or not, but not about whether the gas would be fully stripped from the galactic halo. Therefore in our case, only the mass of stars and gas which dominates the disc potential is sufficient in the consideration of the stripping condition. For the stellar mass, the contributions by both the disc and bulge are probed using the exponential and S\'ersic profiles.

\begin{enumerate}
    \item Gas mass surface density
    
    The VIVA study has presented azimuthally averaged \ion{H}{i} surface densities. However, $\Sigma_{\rm H\scaleto{\rm I}{4pt}}$ averaged across the disc may not well represent the real gas surface density especially for the sample with highly asymmetric \ion{H}{i} morphology which is the main target in this study. Therefore, we re-derived $\Sigma_{\rm H\scaleto{\rm I}{4pt}}$ for eight sectors with a central angle of 45$^\circ$ each which are divided based on the major/minor axis of the stellar disc. A GIPSY\footnote{Groningen Image Processing SYstem, https://www.astro.rug.nl/~gipsy/} task {\tt ELLINT} is used to derive $\Sigma_{\rm g}$ from the VIVA \ion{H}{i} intensity maps.\\
    
    \item Stellar mass surface density
    \label{sec:AP_star}

    In order to derive the stellar mass surface density $(\Sigma_{\rm \star})$, we make use of the Sloan Digital Sky Survey (SDSS) DR12 sky-subtracted $\it {i-}$band images \citep{York2000,Eisenstein2011}. This band is not only a good tracer of low mass stars which generally takes up most of the stellar mass in galaxies but also less affected by dust extinction. Therefore the $\it {i-}$band photometry provides more reliable stellar mass measurement than other shorter wavelength bands \citep{Zibetti2009}. For the entire sample, $\it {i-}$band images with a 5$\sigma$ point source detection limit of 22.3 magnitude are available in the SDSS archive.
    
    We first construct $18 \arcmin\times18\arcmin$ mosaic images centred on each target galaxy keeping a pixel scale of 0.396$\arcsec$ using {\tt SWARP} \citep{Bertin2002}. This is large enough to cover the diffuse stellar light which may be present beyond the virial radius, allowing us to derive a radial profile of the stellar surface density out to a few effective radii. For individual cases, the light from other sources such as foreground stars and nearby galaxies are carefully masked using the IRAF\footnote{Image Reduction and Analysis Facility, http://ast.noao.edu/data/software} task {\tt OBJMASK}, and then manually added. Photometry is conducted using the IRAF task {\tt ELLIPSE}. The centroid of the ellipse is fixed as the galactic centre from the SDSS. The position angle and ellipticity are set to vary with radii. One example is shown in Fig. \ref{fig:Maskimage} using NGC~4522. The fitted rings are shown on the masked $\it {i-}$band image.
    
    We then perform a one-dimensional bulge-disc decomposition to find the best fit for the surface brightness by minimizing $\chi^2$. The model is combined by a S\'ersic profile for a bulge \citep{sersic1963} and an exponential profile for a disc as follows:
    \begin{equation}
    I(R)=I_{\rm e} \exp \Bigg\{ -b_{\rm n} \bigg[\Big(\frac{R}{R_{\rm e}}\Big)^{1/n}-1\bigg] \Bigg\}+I_{\rm d} \exp \Big(-\frac{R}{R_{\rm d}}\Big)
    \end{equation}
    where $R_{\rm e}$ is the effective radius of the S\'ersic profile, $I_{\rm e}$ is the surface brightness at $R_{\rm e}$, n is the S\'ersic index (0-10), $b_{\rm n}=1.9992n-0.3271$ \citep{Capaccioli1989}, $R_{\rm d}$ is the disc scale length, and $I_{\rm d}$ is the surface brightness at $R_{\rm d}$. To take into account the seeing effects, both S\'ersic and exponential profiles are smoothed before the decomposition by performing convolution with a Gaussian function of the seeing of each image. This smooths the unrealistic bumps in the initial result from the fitting process, especially in the outer disc where the signal-to-noise ratio is relatively low. The final, smooth stellar surface density is expressed by the combination of S\'ersic and exponential profiles as shown by an example in Fig. \ref{fig:SBimage}. The black dots are the ellipse fitting result and the green line is $\Sigma_{\rm \star}$ adopted for equation (\ref{eq:GG}) which consists of two contributions, a bulge (red) and a disc (blue).
    
    Finally, we estimate the stellar mass surface density using the mass-to-light ratio for a given $\it {g-i}$ colour \citep{Into2013} adopted from the Extended Virgo Cluster Catalog \citep[EVCC;][]{Kim2014}. The mass-to-light ratio ranges from 1 to 4 in the $\it {g-i}$ colour range of our sample.
    
\end{enumerate}

\begin{figure}
 \includegraphics[width=\columnwidth]{{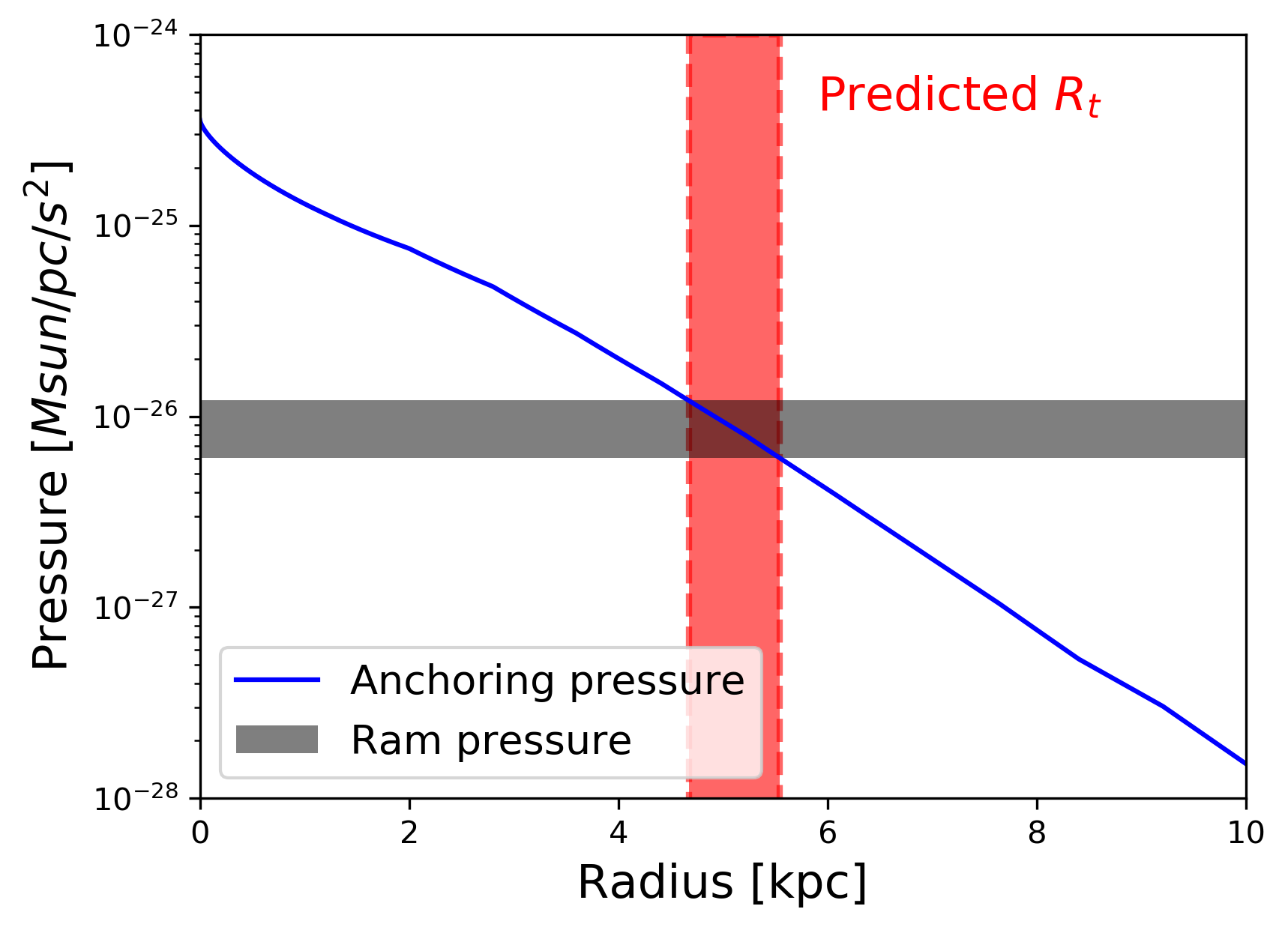}}
 \caption{Illustration of how the predicted truncation radius, $R_{\rm t,pred}$, is measured along one side of the disk. The blue line indicates anchoring pressure estimated as a function of galactic radius along one (out of eight) direction, and the grey shaded area represents the estimation for the surrounding ram pressure at a given clusto-centric distance. The red vertical box indicates the possible $R_{\rm t}$ range of this target based on its stellar and \ion{H}{i} gas surface densities at its location. For each galaxy, the anchoring pressure is measured in all eight segments, so as $R_{\rm t}$, but only one direction is shown as an example.}
 \label{fig:pred_R}
\end{figure}

\subsubsection{Prediction for $R_{\rm t}$} 

We estimate the range of possible $P_{\rm ram}$ at a given clusto-centric distance of each target using equation (\ref{eq:P_ram}). $P_{\rm anch}$ is derived as a function of the galacto-centric radii of each galaxy using equation (\ref{eq:P_anc}). Then the range of predicted $R_{\rm t}$ is found at the position where $P_{\rm ram}$ balances with $P_{\rm anch}$ as shown in Fig. \ref{fig:pred_R}. In this figure, only one direction is shown as an example, but the anchoring pressure and thus $R_{\rm t}$ are measured in all eight segments for each galaxy. As expected, $R_{\rm t}$ is predicted to decrease from the early to post RPS stages with the $R_{\rm t,pred}/R_{\rm 25,opt~B}$ ranges 0.6--1.0 for Class I, 0.3--0.9 for Class II, and 0.3--0.8 for Class III. The mean values are 0.9, 0.6, and 0.5 for Class I, II, and III, respectively.

\begin{table*}
 \begin{threeparttable}
 \caption{Physical quantities of target galaxies from VIVA classified by \citet{Yoon2017}}
 \label{tab:anysymbols}
 \begin{tabular}{|c c c c c c c c c c c c c|}
  \hline
  \hline
  Stage & Galaxy & $d_{\rm M87}$ & $\rho_{\scaleto{\rm ICM}{4pt}}$ & $v_{\rm rel}$ & n & $R_{\rm e}$ & $R_{\rm s}$ & $M_{\star}/L_{i}$ & $M_{\star}$ & $R_{\rm t,obs}$ & $R_{\rm t,pred}$ & $\frac{\Delta R_{\rm t}}{R_{\rm t,obs}}$\\
  & &(deg)&($\rm 10^{-4}\ cm^{-3}$)&($\rm km\ s^{-1}$) & &(kpc)&(kpc)& & ($10^{10}M_{\rm \sun}$) &(kpc)&(kpc)\\
  (1) & (2) & (3) & (4) & (5) & (6) & (7) & (8) & (9) & (10) & (11) & (12) & (13)\\
  \hline
  & N4254 & 3.62 & 0.48 & 1050 - 1485 & 2.94 & 6.55 & 7.05 & 1.26 & 3.23 & 12.39 & 11.35 & 0.083\\
  & N4294\tnote{$\circ$} & 2.54 & 0.79 & 1232 - 1742 & 3.02 & 3.80 & 4.02 & 0.87 & 0.68 & 8.12 & 5.41 & 0.334\tnote{*}\\
  Class I & N4299\tnote{$\circ$} & 2.45 & 0.83 & 1248 - 1765 & 0.16 & 0.11 & 3.06 & 0.65 & 0.17 & 5.67 & 3.27 & 0.423\tnote{*}\\
  : Pre-peak RPS & N4351\tnote{$\circ$} & 1.71 & 1.37 & 1406 - 1989 & 3.82 & 8.00 & 3.87 & 1.10 & 0.27 & 3.46 & 2.61 & 0.246\tnote{*}\\
  (7 galaxies) & N4396\tnote{$\circ$} & 3.49 & 0.50 & 1069 - 1512 & 2.52 & 8.00 & 7.17 & 1.07 & 0.11 & 8.46 & 4.54 & 0.463\tnote{*}\\
  & N4535 & 4.28 & 0.38 & ~~965 - 1365 & 3.05 & 1.06 & 10.69 & 1.47 & 4.90 & 18.38 & 16.88 & 0.082\\
  & N4698 & 5.87 & 0.24 & ~~824 - 1166 & 1.64 & 0.96 & 5.71 & 3.33 & 6.74 & 12.98 & 12.66 & 0.025\\
  \hline
  & N4298 & 3.21 & 0.57 & 1115 - 1577 & 1.61 & 5.05 & 6.44 & 2.32 & 4.04 & 6.85 & 7.08 & 0.033\\
  & N4302 & 3.17 & 0.58 & 1120 - 1584 & 0.22 & 0.99 & 6.51 & 3.49 & 14.56 & 10.33 & 11.58 & 0.121\\
  & N4330 & 2.15 & 0.48 & 1310 - 1853 & 1.29 & 1.57 & 11.14 & 1.73 & 2.64 & 5.62 & 5.12 & 0.088\\
  Class II & N4388 & 1.29 & 2.03 & 1513 - 2140 & 1.36 & 1.64 & 9.00 & 1.98 & 4.68 & 4.99 & 4.53 & 0.092\\
  : Active RPS & N4402 & 1.38 & 1.85 & ~~821 - 1489 & 0.51 & 5.56 & 11.76 & 3.26 & 7.75 & 4.69 & 4.84 & 0.031\\
  (10 galaxies) & N4424\tnote{$\circ$} & 3.11 & 0.59 & 1132 - 1600 & 1.24 & 0.37 & 7.06 & 1.81 & 2.52 & 2.68 & 4.29 & 0.599\tnote{*}\\
  & N4501 & 2.05 & 1.06 & 1330 - 1881 & 2.14 & 1.38 & 5.90 & 3.53 & 31.98 & 9.10 & 9.59 & 0.053\\
  & N4522 & 3.29 & 0.55 & 1101 - 1557 & 1.36 & 1.99 & 9.52 & 1.38 & 1.59 & 4.34 & 4.86 & 0.121\\
  & N4654 & 3.36 & 0.53 & 1090 - 1541 & 1.21 & 0.75 & 6.14 & 1.41 & 4.40 & 9.67 & 9.59 & 0.008\\
  & N4694\tnote{$\circ$} & 4.58 & 0.34 & ~~934 - 1321 & 2.16 & 1.30 & 7.58 & 1.73 & 0.59 & 0.99 & 3.79 & 2.806\tnote{*}\\
  \hline
  & N4064\tnote{$\circ$} & 9.00 & 0.13 & ~666 - 942 & 1.09 & 0.54 & 3.85 & 2.22 & 2.84 & 2.03 & 4.91 & 1.419\tnote{*}\\
  & N4293 & 6.46 & 0.21 & ~~786 - 1112 & 1.83 & 4.49 & 9.46 & 2.37 & 11.73 & - & 4.36 & -\\
  & N4405 & 3.97 & 0.42 & 1003 - 1418 & 1.51 & 1.10 & 8.73 & 2.02 & 1.06 & 2.07 & 2.64 & 0.271\tnote{*}\\
  Class III & N4419\tnote{$\circ$} & 2.83 & 0.68 & 1179 - 1668 & 1.26 & 0.74 & 2.57 & 3.49 & 9.20 & 2.80 & 4.48 & 0.602\tnote{*}\\
  : Post-peak RPS & N4457\tnote{$\circ$} & 8.83 & 0.14 & ~672 - 951 & 1.33 & 1.02 & 4.00 & 2.60 & 4.22 & 4.11 & 5.20 & 0.266\tnote{*}\\
  (10 galaxies) & N4569 & 1.69 & 1.39 & 1411 - 1995 & 10 & 0.72 & 4.25 & 1.93 & 16.55 & 7.62 & 8.24 & 0.083\\
  & N4580\tnote{$\circ$} & 7.24 & 0.18 & ~~743 - 1051 & 0.86 & 0.59 & 12.86 & 2.84 & 2.43 & 2.00 & 2.80 & 0.399\tnote{*}\\
  & N4606 & 2.58 & 0.77 & 1224 - 1732 & 1.06 & 0.49 & 3.86 & 1.98 & 1.94 & 1.49 & 2.49 & 0.668\tnote{*}\\
  & N4607 & 2.64 & 0.74 & 1212 - 1715 & 0.71 & 0.88 & 4.41 & 3.11 & 2.88 & 4.88 & 5.32 & 0.090\\
  & I3392 & 2.66 & 0.74 & 1209 - 1710 & 0.82 & 0.14 & 5.48 & 2.54 & 2.31 & 2.40 & 3.24 & 0.346\tnote{*}\\
  \hline
 \end{tabular}
 \label{tab:targets}
 \begin{tablenotes}
 \item (1) RPS classes adopted from \citet{Yoon2017}, (2) NGC or IC number; those with deviation greater than 1$\sigma$ are indicated with~$^\circ$ (see Fig. \ref{fig:result_all}), (3) projected distance from M87, (4) local ICM density around each target as estimated in Section. \ref{sec:RP}, (5) possible three-dimensional velocity ranges of each galaxy as described in Section. \ref{sec:RP}, (6) S\'ersic index, (7) effective radius of the S\'ersic profile, (8) disc scale length, (9) mass-to-light ratio in $\it {i-}$band as estimated in Section. \ref{sec:AP}, (10) stellar mass derived from the SDSS $\it {i-}$band image and $g-i$ colour as described in Sec. \ref{sec:AP} \ref{sec:AP_star}, (11) shortest $R_{\rm t}$ that can be measured for Class I and II; mean $R_{\rm t}$ through the entire disc for Class III, (12) mean predicted $R_{\rm t}$ for the case with an encounter angle of 45$^\circ$, (13) $|R_{\rm t,obs}-R_{\rm t,pred}|/R_{\rm t,obs}$; those with $\Delta R_{\rm t}/R_{\rm t,obs}$ greater than 3~$\times$~r.m.s of the mean of Class II (0.19) excluding two outliers (i.e., NGC 4424 and NGC 4694) are indicated with~*.
 \end{tablenotes}
 \end{threeparttable}
\end{table*}

\begin{figure*}
 \includegraphics[width=1.95\columnwidth]{{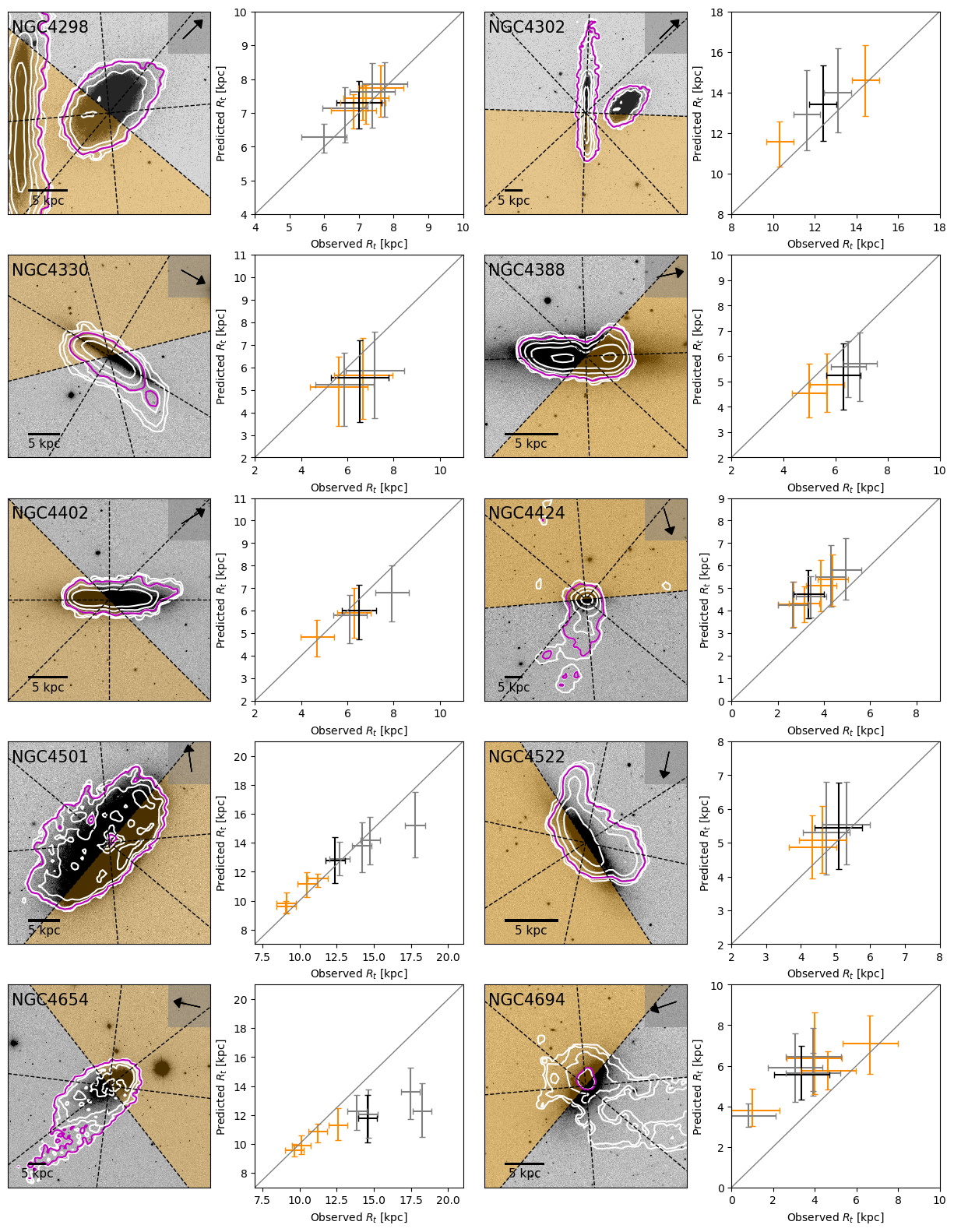}}
 \caption{Comparison of $R_{\rm t}$ (prediction vs. observation) of active RPS cases, i.e., Class II of \citet{Yoon2017}. In the odd columns, the total \ion{H}{i} intensity of each target is shown by white contours overlaid on the SDSS $\it {i-}$band image. The purple line indicates the surface density of $1~M_{\odot}/{\rm pc}^{-2}$ based on which the observational $R_{\rm t}$ is defined. The small arrow on the upper right points the opposite direction of M87. Half of the disc facing directly the ICM wind, i.e., the leading side, is indicated in yellow, whereas the tail side is shaded in grey. The even columns show the comparisons of $R_{\rm t}$ from the calculations based on equation (\ref{eq:GG}) (Section \ref{sec:theory}) and the measurements from observations (Section \ref{sec:observation}). The agreement between the two quantities is shown by a diagonal grey line. The $R_{\rm t}$ of the leading and tail sides are shown in yellow and grey, respectively, and the black points represent the comparison with $R_{\rm t}$ from the azimuthally averaged radial profile. Uncertainties in $R_{\rm t}$ are presented as vertical/horizontal bars.}
 \label{fig:classIIresult}
\end{figure*}
 
\subsection{Observed $R_{\rm t}$ with VLA images}
\label{sec:observation}

We observationally define $R_{\rm t}$ at the radius where the \ion{H}{i} surface density drops to $1~M_{\odot}~{\rm pc}^{-2}$. For the sample undergoing active RPS, more than one $R_{\rm t}$ are required to consider the asymmetric \ion{H}{i} morphology. Therefore we measure $R_{\rm t}$ toward eight directions as described in Section \ref{sec:AP}. The limit of $1~M_{\odot}~{\rm pc}^{-2}$, which is a commonly adopted level to define \ion{H}{i} extents, is also used in this study as it is found to well represent the level where sharp truncations are observed. Consequently, $R_{\rm t}$ to the direction with low surface density tails in a stripped gas trailing region can be underestimated. However, those tail features are not the main topic of interest in this study.

\section{$R_{\rm \texorpdfstring{\MakeLowercase{t}}{}}$: Theory vs. Observation} 
\label{chap:result}

\subsection{Galaxies undergoing active RPS}
\label{sec:ClassII}

The important characteristics of active RPS commonly found in \ion{H}{i} observations are 1) highly disturbed \ion{H}{i} morphology and 2) (regional) truncation within the extent of the stellar disc. The galaxies with these characteristics among the sample are located within a clusto-centric distance of $R_{200}$ (1.55~Mpc for Virgo) at which the ICM surface density estimated by the $\beta$-model is more or less consistent with that observed using X-ray surface brightness. Thus for active RPS cases, i.e., Class II galaxies of \citet{Yoon2017}, ram pressure can be better constrained than the other classes in the early or later stages of RPS, making them ideal targets to verify the GG's formula.

In the first and third columns of Fig. \ref{fig:classIIresult}, the total \ion{H}{i} intensity contours of active RPS galaxies are shown by white contours overlaid on the SDSS $\it {i-}$band images. The thick purple line indicates the \ion{H}{i} surface density of $\rm 1~M_{\odot}~{\rm pc}^{-2}$. As mentioned, the observational $R_{\rm t}$ defined at the \ion{H}{i} surface density of $\rm 1~M_{\odot}~{\rm pc}^{-2}$ is generally a good representative of the \ion{H}{i} extent in the direction where sharp \ion{H}{i} truncation is seen. For each galaxy, $R_{\rm t}$ is measured eight individual segments of an opening angle of 45$^\circ$, i.e., four grey and four yellow sectors. Half of the disc inferred to be facing directly the ICM wind, i.e., the leading side, is shaded in yellow, whereas the tail side is shaded in grey. For reference, the arrow in the upper-right corner of each panel points in the opposite direction of M87. This is a likely ICM wind direction if the galaxy is on the way to the Virgo centre, radially falling in the plane of the sky. It is also the direction in which stripped gas features are expected to roughly point. However, the motion along the observer’s line-of-sight and the galaxy’s recent core-crossing, non radial orbits, or locally non-static ICM can result in some discrepancies in the direction between the arrow and the stripped gas tail. Therefore instead of the location of M87, we infer the effective direction of the ICM wind (i.e., ram pressure) based on the \ion{H}{i} compression and tails.

For each galaxy, $R_{\rm t}$ is measured for eight directions using the \ion{H}{i} radial profiles of eight sectors. For the highly inclined systems ($i>75^\circ$), the \ion{H}{i} extent along the stellar minor axis is comparable with the synthesized beam size $\sim$1 kpc. In such cases, $R_{\rm t}$ along the minor axis is measured to be too large to represent the \ion{H}{i} truncation, and thus we do not use four $R_{\rm t}$’s along the minor axis for highly inclined galaxies in further analysis. The second and fourth columns of Fig. \ref{fig:classIIresult} show the comparisons of $R_{\rm t}$ from the predictions based on (\ref{eq:GG}) (Section \ref{sec:theory}) and the measurements from the observations (Section \ref{sec:observation}). The diagonal grey line is an equal line of the two measurements. The $R_{\rm t}$ of the leading and tail sides are shown in yellow and grey, respectively. The $R_{\rm t}$ from the azimuthally averaged radial profile is shown in black for comparison. The measurement uncertainties in $R_{\rm t}$ are presented as vertical/horizontal bars. The centre of each predicted $R_{\rm t}$ represents the mean of the lower and upper limits for the case with an encounter angle of 45$^\circ$. Meanwhile, the size of each bar shows the possible $R_{\rm t}$ for the entire angle range explored in this study.

For most active RPS cases, the observed $R_{\rm t}$ agrees with the predicted $R_{\rm t}$ within the measurement uncertainties at least for one region of the leading side (yellow colour). 
The true gas truncation radius is likely to be the shortest $R_{\rm t}$ from one of the leading sectors. This approach is similar to how \citet{Gullieuszik2020} defined their $R_{\rm t}$ as the shortest extension of the H$\alpha$ emission along the galaxy major axis.
NGC~4298, NGC~4302, and NGC~4522 generally show a great agreement between $R_{\rm t,obs}$ and $R_{\rm t,pred}$ for all directions. The mean of $\Delta R_{\rm t}/R_{\rm t,obs}$ is $\lesssim$ 0.06 for these three galaxies. On the other hand, half of the active RPS cases show discrepancies on the tail side (i.e., the opposite side of the interface with the ICM wind). They are NGC~4330, NGC~4388, NGC~4402, NGC~4501, and NGC~4654. For this group, the mean $\Delta R_{\rm t}/R_{\rm t,obs}\sim$~0.07 on the leading side, whereas it is 0.13 on the tail side. This implies that the GG's equation predicts a truncation radius reasonably well for a diffuse gas like \ion{H}{i} on the side which is directly affected by the ICM wind. That is, broadly speaking, RPS can be understood as a process of the ICM momentum pushing the ISM from the galactic disc.
Furthermore, this supports the idea that the shortest $R_{\rm t}$ on the leading side is most probable to represent the true gas truncation radius for further similar works.
%\textcolor{red}{However, some galaxies may need caution due to potential bias from their high inclinations.}

On the one hand, 2 out of 10 active RPS cases -- NGC 4424 and NGC 4694 -- show relatively large deviations on a one-to-one line.
NGC~4424 tends to show overall larger offsets between the predicted $R_{\rm t}$ and the observed $R_{\rm t}$ in all directions compared to the others in the same group. This galaxy has a hint of minor merging \citep{Kenney1996, Cortes2006} which may have disturbed the potential, and thus the estimation of its anchoring pressure may not be straightforward. In addition, this galaxy is located close to M49, the centre of a sizable sub-cluster which is in the process of merging to the main Virgo cluster from the south \citep{Gavazzi1999,Mei2007}. In fact, the systematic velocity makes this target more likely to be part of the M49 group \citep{Sancisi1987,Chung2009}. Indeed, the possibility that this galaxy had interacted with the M49's hot halo has been suggested by \citet{Biller2004}. If so, our M87-based estimation must not be suitable for this case. Thus the merging history and membership may result in some discrepancies in the comparison of $R_{\rm t}$ in this case. NGC~4694 also shows somewhat different patterns from the others in the sense that most of the measurements are offset from the grey line. NGC~4694 has a low surface brightness neighbour which is connected through an \ion{H}{i} bridge \citep[VCC 2062;][]{vanDriel1989,Chung2009}. Thus, our simple comparison is unlikely to work properly for this case as well. 

\subsection{Galaxies at pre- and post-peak RPS stages}
\label{sec:result_stages}

\begin{figure}
 \centering\includegraphics[width=\columnwidth]{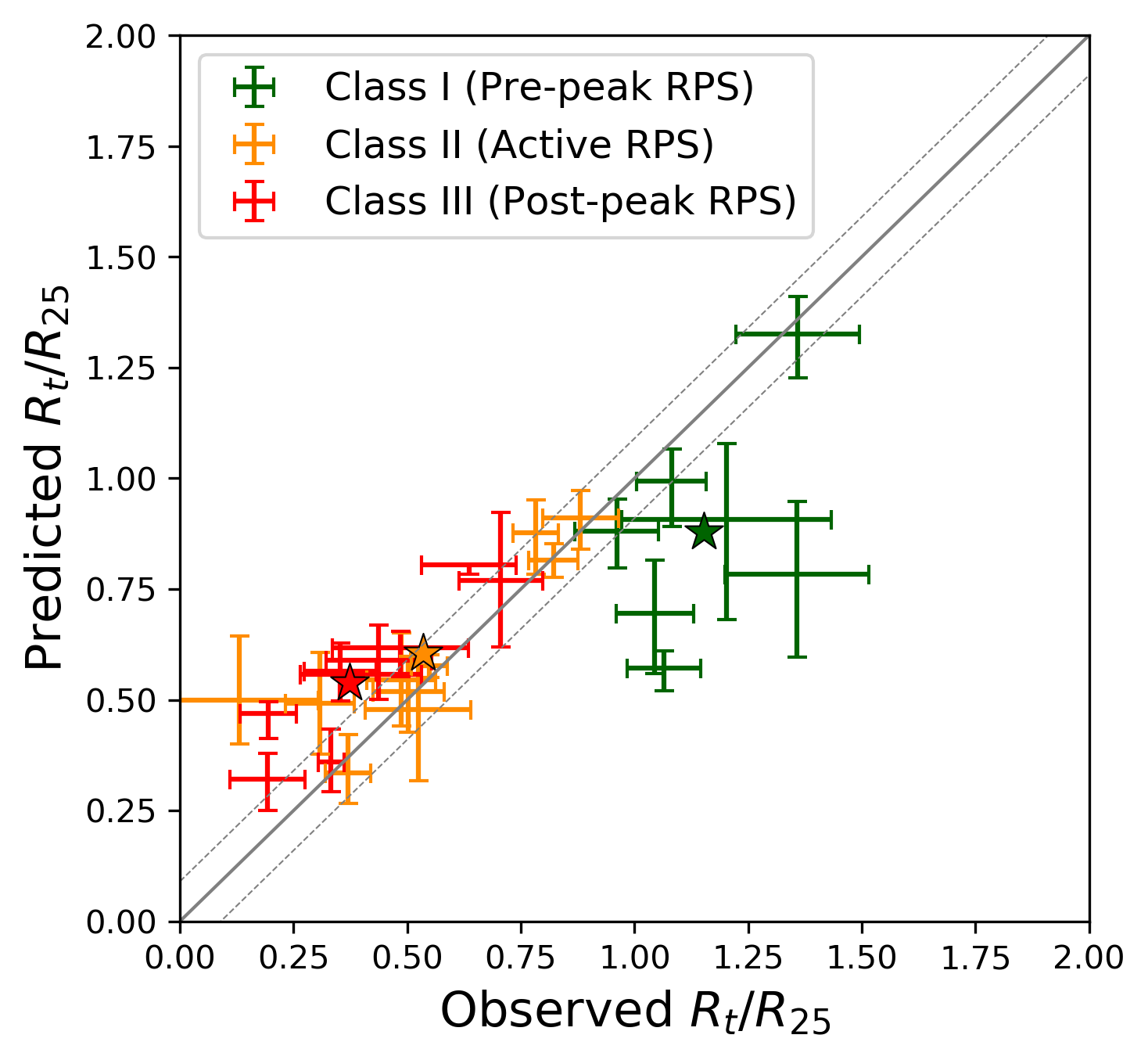}\par
 \caption{Comparisons of $R_{\rm t}$ for pre-peak, active, and post-peak RPS cases in $R_{\rm 25}$, where $R_{\rm 25}$ is the optical size of the semi-major axis measured at 25~mag~$\rm arcsec^{-2}$ in B-band. The diagonal line indicates the equal $R_{\rm t}/R_{\rm 25}$ with $\pm1\sigma$ lines in dashes. The mean $R_{\rm t}$ for each group is marked with a star. Compared to active RPS cases which are offset by 0.07~$R_{\rm 25}$ (yellow), the other two groups are offset by 0.27~$R_{\rm 25}$ for pre-peak RPS cases (green), and 0.16~$R_{\rm 25}$ for post-peak RPS cases (red), respectively.
 }
 \label{fig:result_all}
\end{figure}

In Fig. \ref{fig:result_all}, we compare the samples undergoing various RPS stages. For the pre-peak and active RPS cases (Class I and II of \citet{Yoon2017}), the shortest $R_{\rm t}$ in the leading side (i.e., the opposite side of extended tails) of each galaxy is plotted. Again, the measurements along the minor axis are not considered when $i$>75$^\circ$. For the post-active RPS cases with relatively symmetric \ion{H}{i} morphology, a single $R_{\rm t}$ from the azimuthally averaged radial profile is plotted for each galaxy. Among this group, NGC~4293 is excluded as the \ion{H}{i} surface density does not exceed $\rm 1~M_{\odot}~pc^{-2}$ throughout the disc, and its $R_{\rm t}$ is not defined. 

First of all, it is quite noticeable that $R_{\rm t}$ decreases from pre-peak to active, and then post-peak RPS cases in both predictions and observations. However, the pre- and post-peak RPS groups show somewhat larger deviations from the equal-$R_{\rm t}$ relation compared to the active RPS cases.
On average, the predicted $R_{\rm t}$ of the pre-peak RPS group is smaller than its observationally measured $R_{\rm t}$ by 0.27~$R_{\rm 25}$, whereas the predicted $R_{\rm t}$ is larger than the observed $R_{\rm t}$ by 0.16~$R_{\rm 25}$ for the post-peak RPS group.
That is, the other classes deviates from the one-to-one correlation by a few times on average.
The mean $R_{\rm t}$’s of individual groups are marked with star symbols in the figure.

The pre-peak RPS cases are found with overall larger offsets compared to the other two groups including four cases (57\%) with a noticeably large offset as presented in Table \ref{tab:targets} and Fig. \ref{fig:result_all}. In particular, their observed $R_{\rm t}$ is systematically larger than the predicted $R_{\rm t}$, indicating that their \ion{H}{i} is not stripped as much as expected based on their properties and locations. It means our estimation of the ram pressure for this group is generally overestimated. In fact, most of the sample in this group is located at the clusto-centric distance where no detectable X-ray emission is present (Fig. \ref{fig:xray}), yet we estimate the surrounding ICM density by extrapolating the $\beta$-model. The slope of the modelled ICM density could be less steep than that in reality, which might have resulted in the overestimation of the ICM density for this group, so as the strength of ram pressure. Alternatively, at the stage where the gas starts being pushed strongly by ram pressure yet not stripped, the \ion{H}{i} density across the disk can be overall decreased, proceeding to the next stage (i.e., Class II). This effect may lead to small predicted $R_{\rm t}$ but still a large observed $R_{\rm t}$ for some pre-stripping cases. Indeed, the three extreme outliers in Fig. \ref{fig:result_all} are NGC 4294/9 and NGC 4396 which have been identified with a long one-sided tail, i.e., the sign of quite strong ram pressure among the same group \citep{Chung2007}.

On the one hand, almost 80\% of the post-peak RPS cases show offsets to the opposite, i.e., $R_{\rm t,pred}$ is generally larger than what is actually measured (see Table \ref{tab:targets}). The mean $\Delta R_{\rm t}/R_{\rm t,obs}$ for these outliers is 0.57, which is about seven times the value for the other two post-peak RPS cases, NGC 4569 and NGC 4607. Among them, four cases even deviate by more than 1$\sigma$ from one-to-one line in Fig. \ref{fig:result_all}. This is consistent with the inferred history for this group that they have gone through strong RPS in the past in the higher density region as indicated by the severe \ion{H}{i} truncation, and they have moved out to lower-density regions.

Our results suggest that the understanding of a ram pressure stripped galaxy from the perspective of GG's formula is quite secure when the target is undergoing active RPS in the environment where the surrounding pressure due to the ICM can be well defined. If the galaxy is not close to the peak-pressure or RPS is not a dominant mechanism affecting the target, more caution would be needed to assess the impact of ram pressure on the target.

\subsection{Comparison with the literature}
\label{sec:result_Koppen}

There have been quite a number of challenges to analytically understand GG's relation by comparing ram pressure and anchoring pressure \citep[e.g.][]{Abadi1999,Vollmer2001,Roediger2005,Jachym2007,Steinhauser2016}. Among many previous studies, our work must be most comparable with \citet{Koppen2018} considering that both studies use the same \ion{H}{i} data and estimate pressures in similar ways. However, there are further details for which the two studies approach differently, and the comparison of our results with Köppen's is one way to confirm how broadly GG’s formula can be adopted to interpret the observations in spite of many uncertainties.

In \citet{Koppen2018}, they estimated $P_{\rm loc}$, the expected local ram pressure at a given location of individual galaxies, and $P_{\rm cfg}$, the maximum ram pressure that each galaxy could have experienced (see their Fig. 14 of \citet{Koppen2018}). Their $P_{\rm loc}$ is comparable with our $P_{\rm ram}$ in a sense that both have been calculated assuming a smooth and spherically symmetric ICM distribution based on the best fit $\beta$-model of the Virgo cluster \citep{Schindler1999,Vollmer2009,Vollmer2001} at a given location. However, the two studies use different clusto-centric distance measurements. As in \citet{Jaffe2015}, we first scale the projected distance from M87 by $\pi$/2 to estimate the 3D angle, taking the average of the possible projection angle (i.e., $r = cos\phi~r_{\rm 3D}$ where $0<\phi<\pi\mathbin{/}2$, and $r_{\rm 3D}\sim(\pi\mathbin{/}2)~r$ on average). As stated, we adopted a Virgo distance of 16.5 Mpc to convert $r_{\rm 3D}$ in the sky to a physical distance. Meanwhile, \citet{Koppen2018} used the projected distance from M87 converted with a uniform distance of 17 Mpc for most of the sample, and adopted the stellar Tully-Fisher distance for 11 galaxies from \cite{Cortes2008}. Regarding the galaxy’s velocity, whereas we use a velocity range for each system from the Keplerian velocity to the escape velocity, \citet{Koppen2018} used only the escape velocity and even this value may differ from ours depending on the assumed Virgo potential.

\begin{figure}
 \centering\includegraphics[width=\columnwidth]{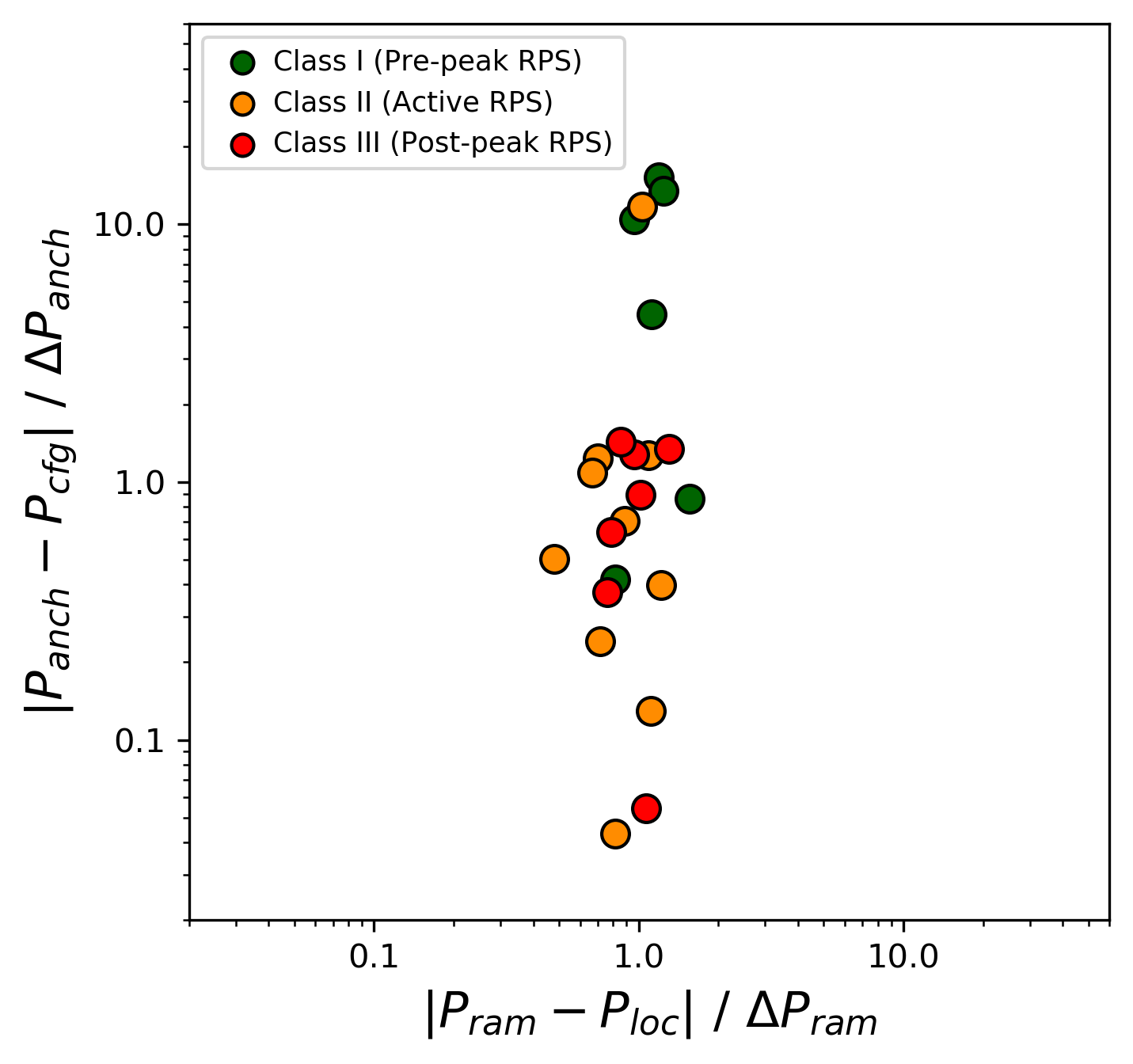}\par
 \caption{Differences between our pressure measurements, $P_{\rm ram}$ and $P_{\rm anch}$, and those of \citet{Koppen2018}, $P_{\rm loc}$ and $P_{\rm cfg}$, which are normalized by our measurement uncertainties, $\Delta P_{\rm ram}$ and $\Delta P_{\rm loc}$, respectively. The ram pressures show good agreement in general, whereas the anchoring pressures show a rather wide range of differences with four extreme outliers.}
 \label{fig:Koppen}
\end{figure}

The x-axis of Fig. \ref{fig:Koppen} shows the difference between our $P_{\rm ram}$ and $P_{\rm loc}$ of \citet{Koppen2018} which is normalized by our measurement uncertainty of $P_{\rm ram}$ (i.e., $|P_{\rm ram}-P_{\rm loc}|/\Delta P_{\rm ram}$). $\Delta P_{\rm ram}$ has been evaluated by considering the potential encounter angle from 0$^\circ$ to 60$^\circ$, and the relative velocity ranges from Keplerian velocity to escape velocity. In spite of the differences in distance and velocity, $|P_{\rm ram}-P_{\rm loc}|$ is comparable with $\Delta P_{\rm ram}$ for most of the cases. Particularly for the active stripping cases which are the main targets of this study, the mean $|P_{\rm ram}-P_{\rm loc}|/\Delta P_{\rm ram}$ is $\sim$ 0.84, and the ram pressure from the two studies is generally in good agreement.

Meanwhile, $P_{\rm cfg}$ of \citet{Koppen2018} can be compared with our $P_{\rm anch}$ although there are some important differences. \citet{Koppen2018} estimated the maximum ram pressure that the galaxy could have experienced to be observed with the current \ion{H}{i} radius under its centrifugal force. For the pre-stripping \ion{H}{i} content, they assumed that the gas surface density follows a Miyamoto-Nagai profile \citep{Miyamoto1975} with constants satisfying the initial \ion{H}{i} mass and initial outer \ion{H}{i} radius. On the other hand, our anchoring pressure is derived as a function of radii for each galaxy using the radial stellar and gas surface densities based on the SDSS and VIVA observations. In addition, \citet{Koppen2018} took the azimuthally averaged \ion{H}{i} radius from \citet{Chung2009} as a stripping radius whereas we make considerations of the \ion{H}{i} asymmetry using the isophotal \ion{H}{i} radii measured along 8 directions which is particularly important for the active stripping cases (i.e., the main targets of this study), and some early stripping cases.

The y-axis of Fig. 8 shows the difference between our $P_{\rm anch}$ and $P_{\rm cfg}$ of \citet{Koppen2018} which is normalized by our measurement uncertainty of $P_{\rm anch}$ (i.e., $|P_{\rm anch}-P_{\rm cfg}|/\Delta P_{\rm anch}$). $\Delta P_{\rm anch}$ has been evaluated considering the spatial resolution of the VIVA data, i.e, assuming the minimum/maximum $R_{\rm t}~\pm\approx15\arcsec$. Compared to $P_{\rm ram}$, $|P_{\rm anch}-P_{\rm cfg}|/\Delta P_{\rm anch}$ ranges quite broadly from very small values for which the two studies are extremely well consistent to quite large values for which the difference between the two studies is much bigger than our measurement uncertainties. All four outliers are the cases showing significant \ion{H}{i} asymmetry. Among these, three galaxies (NGC 4254, 4294, and 4535) are the ones that do not show \ion{H}{i} truncation within the extent of the stellar disc yet the \ion{H}{i} is quite asymmetric (i.e., early stripping cases). The fourth outlier is NGC 4522 that is undergoing active stripping and shows a very disturbed \ion{H}{i} distribution. This comparison thus shows that the consideration of gas asymmetry can make quite a huge difference in the estimation of anchoring pressure, especially for the cases with a \ion{H}{i} disc that is more extended than the stellar disc along all or some directions. However, for most of the active stripping cases which are severely stripping within the extent of the stellar disc, the anchoring pressure from the two studies is in quite good agreement, showing discrepancies that is comparable with our measurement uncertainty or less.

To summarize, in spite of subtle differences in estimating both ram pressure and anchoring pressure, GG's relation seems to hold for the cases which show clear evidence of \ion{H}{i} truncation within the extent of the stellar disc. For those with highly asymmetric gas distribution yet not stripped severely, extra caution may be needed in the interpretation as the anchoring pressure across the disc may significantly vary.

\section{Discussion}

\subsection{Re-assessment of $R_{\rm t}$}
\label{sec:pre-strip}

\begin{figure}
 \centering\includegraphics[width=\columnwidth]{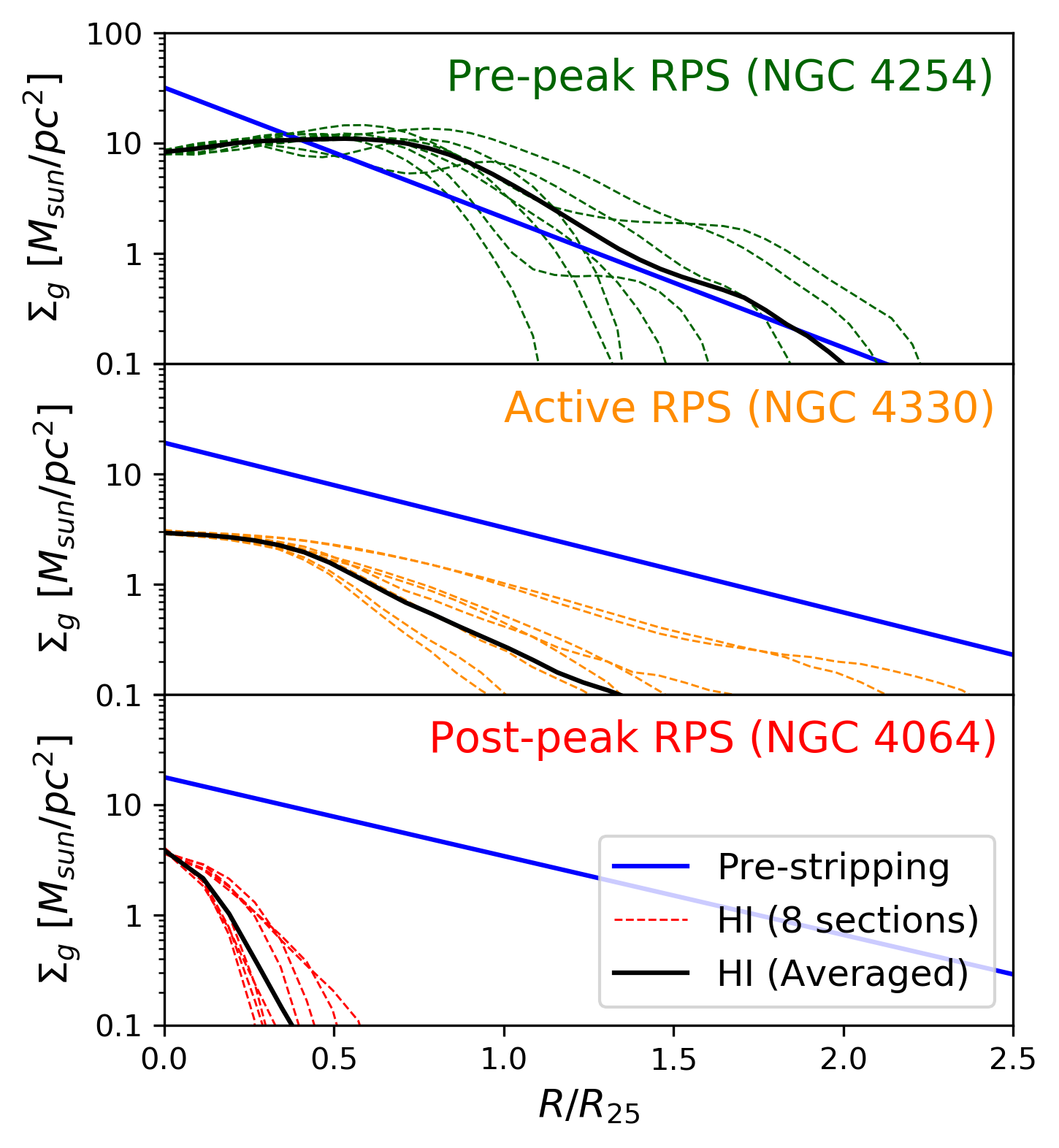}\par
 \caption{Comparison of the radial $\Sigma_{\rm g}$ from the model with the observed \ion{H}{i} surface density of our targets by RPS stages. The solid black line in each panel is the azimuthally averaged $\Sigma_{\rm H\scaleto{\rm I}{4pt}}$ of each example (current), and the solid blue line represents the model (before stripping). The dashed lines indicate the observed \ion{H}{i} surface densities along eight individual directions. The disc of the early RPS group (top panel of Fig. \ref{fig:pre-stripping_HI}), which is not severely affected, more or less agrees with the model. This implies that the model is indeed a good representative of pre-stripping $\Sigma_{\rm g}$. For the active and post-peak RPS groups, the model of pre-stripping shows quite large differences, indicating that our initial calculation of anchoring pressure could have been underestimated, so as $R_{\rm t}$}
 \label{fig:pre-stripping_HI}
\end{figure}

To estimate the expected $R_{\rm t}$, we adopted the $\Sigma_{\rm g}$ from the observed \ion{H}{i} surface radial profiles regardless of the galaxy's RPS status. However, the gas column density before stripping is likely higher, and the mean gas density and size must be continuously decreasing during the RPS process as seen in Fig. \ref{fig:pre-stripping_HI} \citep[see also][]{Cayatte1994}. That is, the actual density/extent right before the galaxy got stripped and reaches the current status must be slightly higher/larger than what we assumed. In this section, we therefore test how the gas conditions before stripping would change our results. Since it is highly uncertain by how far the gas was denser right before a galaxy is observed to be as now, we do the test adopting an exponential profile as \citet{Jaffe2018} do for gas-rich disky galaxies.

%We adopt the methodology of \citet{Jaffe2018} to estimate $\Sigma_{\rm g}$ before stripping. 
The gas has an exponential profile described as follows:	
\begin{equation}
    \Sigma_{\rm d,gas}(R)=\Sigma_{\rm 0,gas} \exp \Big(-\frac{R}{R_{\rm s,gas}}\Big)
\end{equation}
where $R_{\rm s,gas}$ is the gas scale length, $\Sigma_{\rm 0,gas}(=M_{\rm gas}/2\pi\cdot{R_{\rm s,gas}}^2$) is the central gas surface density, and $M_{\rm gas}$ is the total gas mass. The gas scale length, $R_{\rm s,gas}$, is assumed to be 1.7 times larger than the stellar disc based on the non-\ion{H}{i}-deficient Virgo spirals \citep{Cayatte1994}. The cool gas (\ion{H}{i} and H$_2$) mass fraction to the stellar mass is adopted from the known relation for disc-dominated galaxies with B/T$<$0.4 as follows \citep{Popping2014,Gullieuszik2020}:
\begin{equation}
    f_{\rm gas}=\frac{M_{\rm g}}{M_{\rm \star}}=0.158\Big(\log \frac{M_{\rm \star}}{M_{\rm \odot}}\Big)^2-3.548 \log\frac{M_{\rm \star}}{M_{\rm \odot}} +19.964
\end{equation}
where $M_{\rm gas}$ and $M_{\rm \star}$ are the masses of cool gas and stars, respectively. In Fig. \ref{fig:pre-stripping_HI}, the predicted radial $\Sigma_{\rm g}$ from the model is compared with the observed \ion{H}{i} surface density by RPS stages.

\begin{figure}
 \centering\includegraphics[width=\columnwidth]{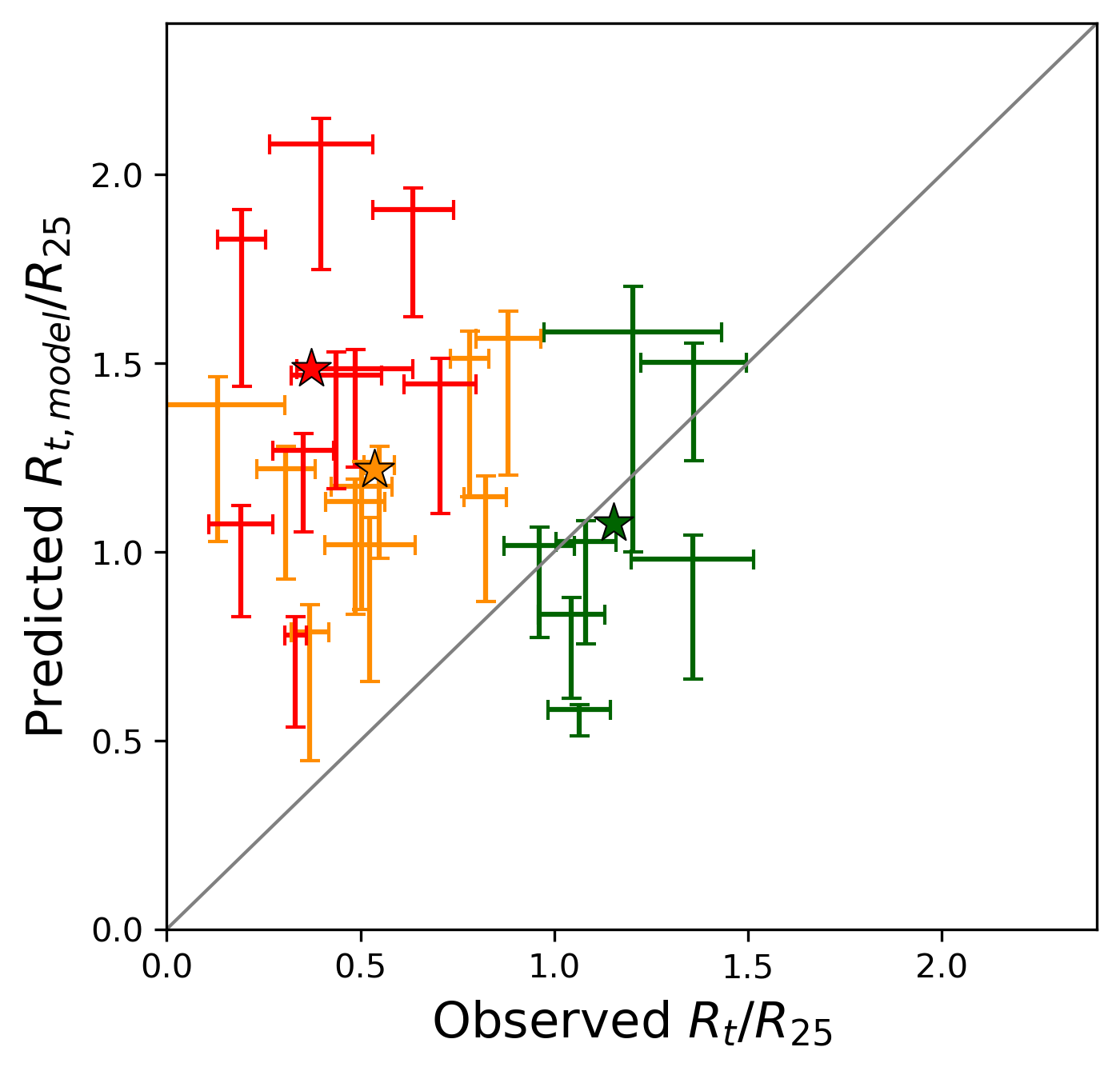}\par
 \caption{Comparisons of the re-estimated, expected $R_{\rm t}$ and the observed $R_{\rm t}$ in \ion{H}{i}. Other details are the same as Fig. \ref{fig:result_all}. Since the gas surface density before stripping is highly uncertain, here we adopted an exponential profile as \citet{Jaffe2018} as the pre-stripping gas surface density, $\Sigma_{\rm g}$, which leads the expected $R_{\rm t}$ to overall shifted up.}
 \label{fig:pre-stripping_Rt}
\end{figure}

As expected from the concept of GG, Fig. \ref{fig:pre-stripping_HI} clearly shows that the stripping proceeds from outside to inside on a gas disc. In addition, it is also intriguing to see a dramatic decrease of the gas surface density in the inner part ($\lesssim 0.5R/R_{\rm 25}$). This evidently shows that the inner gas disc can be severely affected by the ram pressure in spite of a large potential and higher density as reported in both simulations \citep[e.g.][]{Vollmer2001,Steinhauser2016,Lee2020} and observations \citep[e.g.][]{Wang2014,Mok2017,Lee2017}. It should be noted that molecular gas is not included in the observed gas surface density in this comparison. If the molecular phase is added, the gas surface density can be higher by a factor of a few tens in the central region \citep[e.g.][]{Bigiel2008}. However, the molecular gas density rapidly drops with increasing radii, and around $\sim$0.5~$R_{\rm 25}$, two phases become comparable and the atomic phase takes over at larger radii. Therefore, at $R_{\rm t}$ of active and post-peak RPS cases ($\sim$0.54~$R_{\rm 25}$ and $\sim$0.37~$R_{\rm 25}$), the total surface density can be several times higher than that of atomic gas alone, which will lead to a few times higher anchoring pressure than what we estimated. However, a few times higher gas surface density results in the increase of $R_{\rm t}$ only by a kilo-parsec or so as seen in Fig. \ref{fig:pred_R}. This is smaller than the usual uncertainties in $R_{\rm t}$, and thus the presence of molecular gas is thought to hardly affect our results such as Fig. \ref{fig:result_all} and Fig. \ref{fig:pre-stripping_Rt}.

%\textcolor{blue}{When the molecular phase is taken into account, the difference would be actually smaller. Therefore this comparison between the model and the observed $\Sigma_{\rm H\scaleto{\rm I}{4pt}}$ provides the upper limit on the difference between the two.}
%\textcolor{red}{In addition, the absence of molecular gas content for the observed gas surface density seems to have little effect on measuring truncation radius for most samples in our study. According to Fig. 6 of \citet{Lee2017}, an extent of CO seems to be smaller than that of HI for some of our active RPS cases -- NGC 4330, NGC 4402, and NGC 4522, while an extent of HI and CO are comparable for the post-peak RPS case, NGC 4569. This implies that the potential from molecular gas to the anchoring pressure can be ignored in most pre-peak and active RPS cases, but there may be a possible bias in some post-peak RPS cases to the results.}

In Fig. \ref{fig:pre-stripping_Rt}, the comparisons of the re-estimated expected $R_{\rm t}$ and the \ion{H}{i} observed $R_{\rm t}$ are presented. By using the pre-stripping $\Sigma_{\rm g}$, the expected $R_{\rm t}$ is overall shifted up when compared with Fig. \ref{fig:result_all}. As discussed in Section \ref{sec:result_stages}, $\Sigma_{\rm g}$ of some pre-stripping cases is expected to be lowered. Thus this group agrees better in this relation in Fig. \ref{fig:pre-stripping_Rt} compared with in Fig. \ref{fig:result_all}. However, this assumption (i.e., $\Sigma_{\rm g}$ follows that of the model) is another extreme condition since $\Sigma_{\rm g}$ at the time of stripping is more likely to be somewhere between the model and the currently observed $\Sigma_{\rm H\scaleto{\rm I}{4pt}}$. Indeed, the expected $R_{\rm t}$ based on the model does not seem to be too realistic. The ranges of expected $R_{\rm t}$ are comparable regardless of the RPS stage unlike in Fig. \ref{fig:result_all} where $R_{\rm t}$ is found to decrease from early, active to past RPS stages as the clusto-centric distance of each group also decreases overall. 

These indicate that using $\Sigma_{\rm g}$ from \ion{H}{i} imaging data for the relevant estimation should be reasonable. Although the observed \ion{H}{i} surface density may result in the underestimation of the predicted $R_{\rm t}$, it must still be closer to the actual $\Sigma_{\rm g}$ at the time of stripping than the model of a gas disc that is intact at all.

\begin{figure}
 \centering\includegraphics[width=\columnwidth]{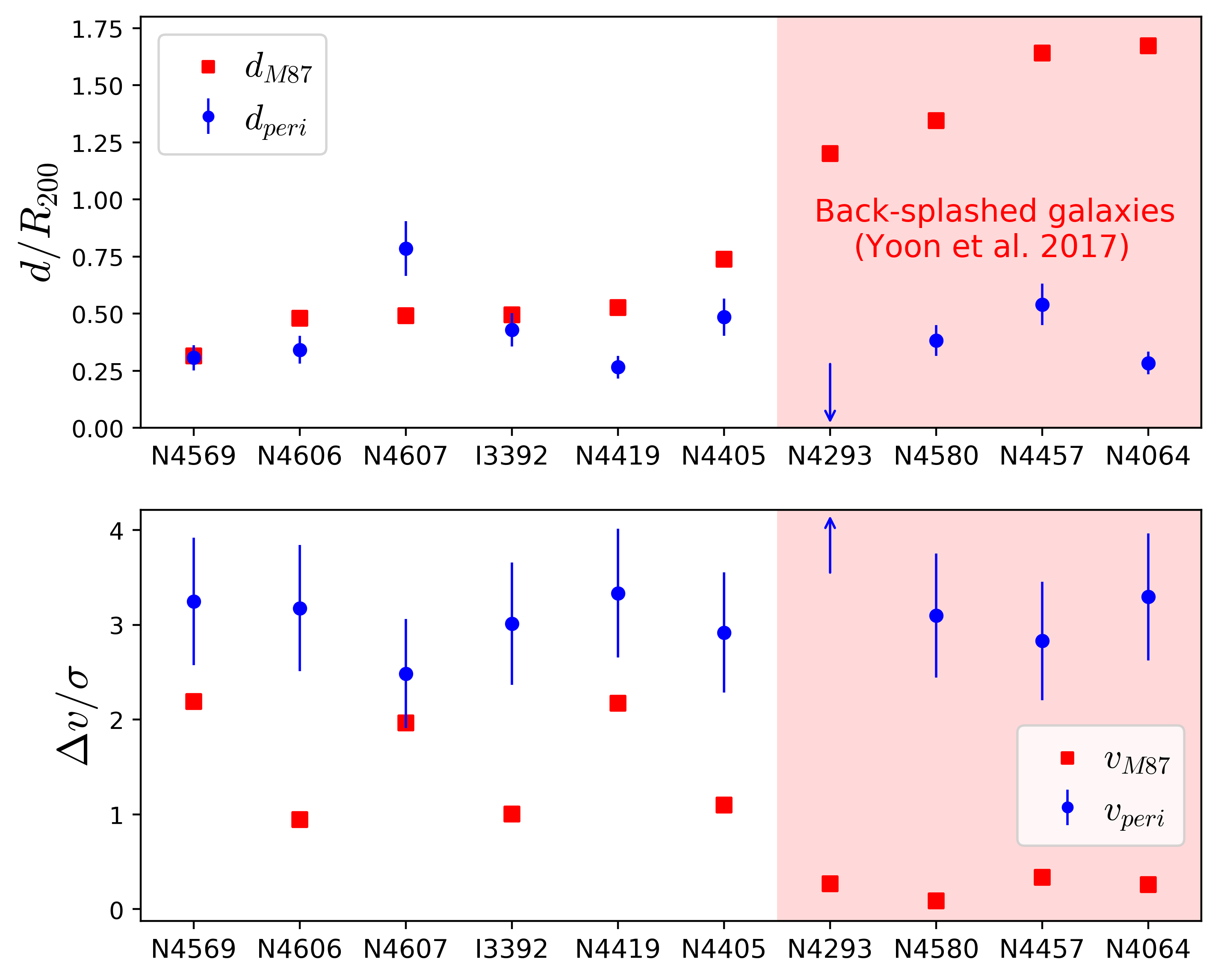}\par
 \caption{(Top) The blue dots represents the possible periapsis of post-peak RPS galaxies computed based on the ram pressure that is required to produce the observed \ion{H}{i} extent ($d_{\rm peri}$). The red squares represent the clusto-centric distances of the sample at their current locations ($d_{\rm M87}$). Both are normalized by $R_{\rm 200}$ \citep[=1.55 Mpc;][]{McLaughlin1999,Ferrarese2012}. The red shaded region indicates the back-splashed cases identified by \citet{Yoon2017}. Galaxies are shown in the order of increasing $d_{\rm M87}$ from left to right. (Bottom) Comparison between the possible maximum velocity relative to the cluster centre, i.e., the velocity at the pericentre, and the observed line-of-sight velocity relative to the cluster centre for the same galaxies. The colour scheme is identical to that of the upper panel and the velocities are normalized by the cluster dispersion of $\sigma_{\rm cl}$ = 593 km/s \citep{Mei2007}.
 }
 \label{fig:pericenter}
\end{figure}

\subsection{Re-assessment of $P_{\rm ram}$ for post-peak RPS galaxies}

% Periapsis expectation of Class III
\label{sec:pericenter}

In the estimation of the ram pressure experienced by each galaxy, $\rho_{\scaleto{\rm ICM}{4pt}}$ and $v_{\rm rel}$ have been inferred based on the projected clusto-centric distance of the target. However, the galaxies that have gone through strong RPS a while ago, i.e., Class III of \citet{Yoon2017}, may have lost the gas somewhere closer to the cluster centre than where they are currently located. Therefore, our estimation of ram pressure for those galaxies may not be sufficient to explain how deficient and truncated their gas discs are. As seen in Fig. \ref{fig:result_all}, the expected $R_{\rm t}$ of most post-peak RPS cases are indeed found to be larger than the observed $R_{\rm t}$.

In order to probe how deep the post-peak RPS cases have been in the cluster, we estimate the closest approach to the cluster centre (periapsis) which potentially better explains their \ion{H}{i} properties. Assuming that the \ion{H}{i} disc of each galaxy has reached its observed size around its periapsis, we first find the ram pressure which balances with the anchoring pressure at the observed \ion{H}{i} radius. We then find the clusto-centric distance which matches with this ram pressure.

Fig. \ref{fig:pericenter} shows the estimates for the periapsis of the post-peak RPS cases. The upper panel compares the possible periapsis ($d_{\rm peri}$) with the projected clusto-centric distance ($d_{\rm M87}$) of the current location. From left to right, galaxies are shown in the order of increasing $d_{\rm M87}$. Except for one galaxy (NGC~4607), all the post-peak RPS cases are found to have been closer than $\sim$0.5$R_{\rm 200}$ from M87, suggesting that they experienced stronger ram pressure in the past. In addition, those suggested as back-splashed galaxies by \citet{Yoon2017} indeed show the largest discrepancies between $d_{\rm peri}$ and $d_{\rm M87}$ (shaded in red). In the case of NGC~4607, it is highly inclined ($i>75^\circ$) and the vertical scale height of its \ion{H}{i} disc is comparable to the \ion{H}{i} resolution, and the derived radial profile, hence $R_{\rm t}$ is highly uncertain.

The lower panel of Fig. \ref{fig:pericenter} shows the comparison of the possible $v_{\rm rel}$ to the cluster centre at the periapsis ($v_{\rm peri}$) and the \ion{H}{i} observed line-of-sight velocity to the cluster mean. 
We first calculated the potential velocity that these galaxies might have had in the past at the periapsis. For this, we adopt $P_{\rm ram}$ which balances with $P_{\rm anch}$ \Big(i.e., $v=\sqrt{\frac{P_{\rm ram}}{\rho_{\scaleto{\rm ICM}{4pt}}(d_{\rm peri})}}=\sqrt{\frac{P_{\rm anch}(R_{\rm t,obs})}{\rho_{\scaleto{\rm ICM}{4pt}}(d_{\rm peri})}}$\Big).
In addition, we assume that the actual velocity can range between this Keplerian velocity and the escape velocity at the periapsis of each galaxy, which is shown in Fig. \ref{fig:pericenter}. All the galaxies in this group are inferred to have had higher velocities than the currently observed line-of-sight velocities. In particular, the back-splashed cases are thought to have been affected by ram pressure that is much stronger than what they are experiencing in their current location by a factor of up to $\sim$100 as inferred from $d_{\rm peri}$ and $v_{\rm peri}$.

\section{Summary \& Conclusions}

In this study, we attempted to verify our understanding of the RPS based on the GG's relation by comparing the observed \ion{H}{i} radius of the VIVA sample in different RPS stages (pre-peak, active, and post-peak RPS) with the predicted gas extent under the given conditions of individual galaxies. The important lessons from this study can be summarized as follows.

\begin{enumerate}
    \item For active RPS galaxies, the GG's condition works reasonably well (Fig. \ref{fig:classIIresult}).
    These galaxies generally show good agreement between the theory and the observation, particularly on the leading side, which is directly affected by the wind.\\
    \item Galaxies in the pre-/post-peak RPS stage tend to show a larger/smaller observed truncation radius than what is predicted (Fig. \ref{fig:result_all}).
    We conclude that if the ram pressure is not a dominant mechanism for the gas stripping, or if the galaxy is not in an active stripping phase, the impacts of other factors need to be carefully considered.\\
    \item We attempted to verify the results (i) and (ii) using the theoretical gas surface density before stripping other than the \ion{H}{i} surface density. However, the predicted gas extent shows comparable ranges for the entire sample regardless of the RPS stage which is unrealistic (Fig. \ref{fig:pre-stripping_Rt}). This indicates that the observed \ion{H}{i} gas surface density provides more reasonable estimations even though it must be still somewhat lower than the density right before stripping.
    \\
    \item We estimated the periapsis and the velocity at the periapsis of post stripping galaxies using the observed truncation radius (Fig. \ref{fig:pericenter}).
    The galaxies that show large deviations between the periapsis and the clusto-centric distance are inferred to have been deeper inside the cluster in the past. This is consistent with \citet{Yoon2017}'s claim based on the phase-space analysis that they are back splashed cases.
\end{enumerate}

The GG's equation is rather simplified and the discs of real galaxies are not infinitely thin as it assumes \citep[see discussion of caveats in][]{Jaffe2018}. However, the GG's equation provide good predictions for the gas extent of galaxies undergoing RPS. This is also confirmed by the comparison with the work of \citet{Koppen2018}, which used the same VIVA \ion{H}{i} data but different ram pressure and galaxy parameters from ours.
This indicates that the process of gas stripping itself can still be understood as the ICM momentum pushing out the ISM from a galaxy.
This study can be extended using RPS galaxies in other clusters or galaxies in a group environment. Advances in future telescopes, such as SKA, are anticipated to make these extensions possible with unprecedented sensitivity and resolution.

\section*{Acknowledgements}

We are grateful to the anonymous referee for useful comments and suggestions which helped to improve the manuscript tremendously. AC and SL acknowledge support by the National Research Foundation of Korea (NRF), Grant No. 2018R1D1A1B07048314, 2022R1A2C1002982, and 2022R1A6A1A03053472. YKS acknowledges support from the National Research Foundation of Korea (NRF) grant funded by the Ministry of Science and ICT (NRF-2019R1C1C1010279). HY acknowledges support from the Australian Research Council Centre of Excellence for All Sky Astrophysics in 3 Dimensions (ASTRO 3D), through Project No. CE170100013. YJ acknowledges financial support from ANID BASAL Project No. FB210003 and FONDECYT Iniciaci\'on 2018 No. 11180558.

\section*{Data Availability}
The data underlying this article will be shared on reasonable request to the corresponding author.

%%%%%%%%%%%%%%%%%%%%%%%%%%%%%%%%%%%%%%%%%%%%%%%%%%



%\bibliographystyle{mnras}

\bibliographystyle{mnras}
\bibliography{main_ac}

\end{document}